\documentclass[acmsmall, screen]{acmart}
\AtBeginDocument{%
  }

\setcopyright{acmlicensed}
\copyrightyear{2018}
\acmYear{2018}
\acmDOI{XXXXXXX.XXXXXXX}
\acmConference[Conference acronym 'XX]{Make sure to enter the correct
  conference title from your rights confirmation email}{June 03--05,
  2018}{Woodstock, NY}
\acmISBN{978-1-4503-XXXX-X/2018/06}

\usepackage{enumitem}
\usepackage{booktabs}
\usepackage{algorithm}
\usepackage{algpseudocode}
\usepackage{amsmath}
\usepackage{tabularx}
\usepackage{makecell}
\usepackage{graphicx}
\usepackage[most]{tcolorbox}
\usepackage{graphicx}
\usepackage{subcaption}
\usepackage{threeparttable}
\usepackage{wrapfig}

\newtcolorbox{findingbox}{
    enhanced,
    colback=gray!5,
    frame hidden,
    boxrule=0pt,
    borderline west={4pt}{0pt}{gray!80}, 
    sharp corners,
    left=6pt,
    right=2pt,
    top=2pt,
    bottom=2pt,
    fontupper=\relax,
    parbox=false
}

\newcommand{\Code}[1]{\begin{small}\texttt{#1}\end{small}}

\newcommand{\id}[2]{(#1, Post \##2)}

\begin{document}

\title{Understanding Bugs in Template Engine-Based Applications: Symptoms, Root Causes, and Fix Patterns}

\author{Kai Gao}
\orcid{0000-0002-0942-7890}
\affiliation{%
 \institution{University of Science and Technology Beijing}
 \city{Beijing}
 \country{China}}
\email{kai.gao@ustb.edu.cn}

\author{Yu Sun}
\affiliation{%
 \institution{University of Science and Technology Beijing}
 \city{Beijing}
 \country{China}}
\email{yusun@xs.ustb.edu.cn}

\author{Chang-ai Sun}
\orcid{0000-0003-3696-6176}
\affiliation{%
 \institution{University of Science and Technology Beijing}
 \city{Beijing}
 \country{China}}
\email{casun@ustb.edu.cn}

\begin{abstract}
Template engines are indispensable components in modern software ecosystems, enabling the generation of structured documents and scripts across domains such as web development, Infrastructure as Code, and data engineering. 
However, the unique architectural characteristics of template engine-based applications (i.e., TE applications), including multi-language composition, opaque data flow, deferred validation, and complex integration, pose significant challenges for diagnosing and resolving bugs in TE applications. 
While prior research has primarily focused on template engine security, bugs in TE applications remain under-investigated. 
To bridge this gap, we present the first comprehensive study of TE application bugs. 
By analyzing 1,004 application bugs across 15 template engines in five programming languages, we identify the symptoms and root causes of TE application bugs and common patterns to fix them. 
Our findings reveal that \textit{Abnormal Rendering Result} (e.g., unexpected or blank output) is the most prevalent symptom (48.61\%), often manifesting as silent failures that are difficult to diagnose. 
We identify 17 root causes, with \textit{Syntax Misuse}, \textit{Mismatched Data Context}, and \textit{Incompatible Integration} as the dominant categories. 
Furthermore, we find that while 67.92\% of the bugs are fixed within the template, over 20\% require modifications in the host-side logic to resolve data context issues. 
Based on these findings, we derive actionable implications for tool designers, practitioners, and researchers. 
To demonstrate the practical utility of our findings, we further develop two prototype tools for the Jinja engine to facilitate the development and debugging of TE applications. 
\end{abstract}



\received{20 February 2007}
\received[revised]{12 March 2009}
\received[accepted]{5 June 2009}

\maketitle

\section{Introduction}\label{s: intro}
The separation of business logic from the presentation layer is a fundamental software engineering principle aimed at improving flexibility and facilitating team collaboration~\cite{10.1145/988672.988703}. 
This principle underpins the architecture of numerous high-impact systems, such as the Python Package Index (PyPI)~\cite{PyPIArchitecture}. 
To facilitate this separation, template engines have emerged as vital components, which are specialized software designed to combine dynamic data with static boilerplate (i.e., templates) to generate structured documents~\cite{wikipediaTemplateProcessor}, such as HTML documents, YAML config files, and SQL scripts. 
These engines provide templating languages, allowing users to declare placeholders, execute conditional logic and loops, and perform data transformations within templates. 
Nowadays, template engines have been deeply integrated into a wide array of popular and critical ecosystems, including web frameworks (e.g., Django~\cite{Django} and Spring~\cite{spring}), Infrastructure as Code (IaC) tools (e.g., Ansible~\cite{ansible}), and data orchestration platforms (e.g., Airflow~\cite{airflow} and dbt~\cite{dbt}). 

\begin{figure}
    \centering
    \includegraphics[width=\linewidth]{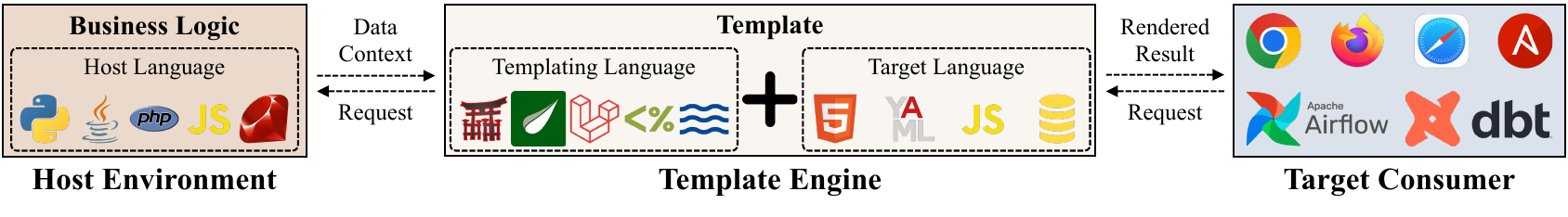}
    \caption{An architectural overview of TE applications.}
    \label{fig: application-overview}
\end{figure}

\begin{figure}
    \centering
    \includegraphics[width=\linewidth]{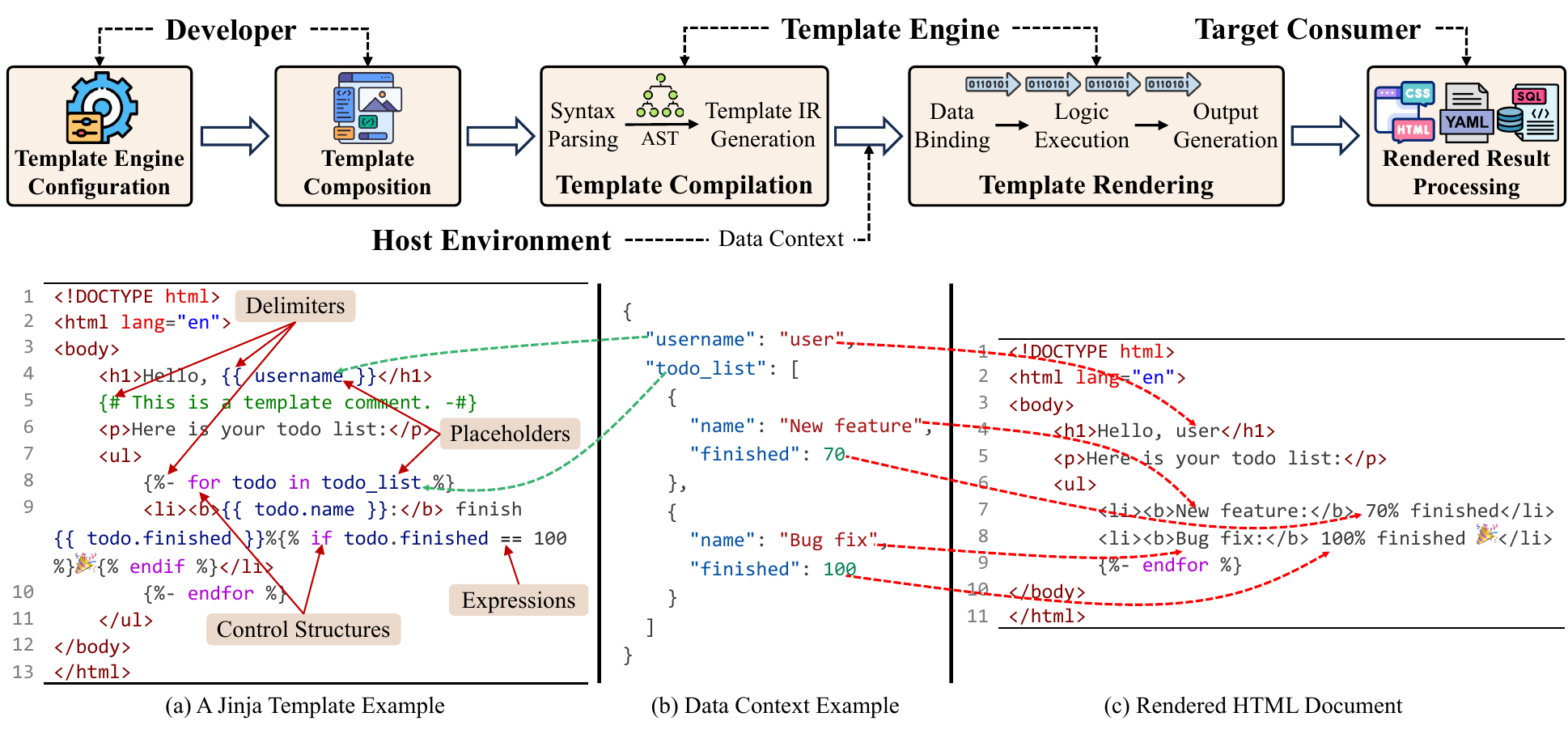}
    \caption{The development workflow of TE applications. The template example (a) is written in the templating language provided by the Jinja engine. Given the data context (b) provided by the host environment, Jinja renders this template into an HTML document (c), which is further processed by the target consumer (e.g., a browser) for final visualization.}
    \label{fig: development-worflow}
\end{figure}

However, template engine-based applications (\textit{TE applications} for short) exhibit several unique complexities, which pose challenges to diagnosing and fixing bugs in them: 
\begin{itemize}[leftmargin=*, topsep=0pt]
    \item \textbf{Multi-Language Composition}. TE applications are inherently multilingual. 
    As illustrated in Figure~\ref{fig: application-overview}, they typically involve a \textit{host language} for business logic (e.g., Python), a \textit{templating language} for dynamic rendering (e.g., Jinja), and a \textit{target language} for the final output (e.g., HTML or YAML). 
    Specifically, templates are usually polyglot artifacts~\cite{Polyglot} that interweave templating directives and target-language boilerplate (e.g., Figure~\ref{fig: development-worflow} (a)). 
    This multilingual nature necessitates developers to navigate the syntax nuances across languages~\cite{7476675, 10413900} and often invalidates existing static analysis tools, which are typically optimized for single-language settings~\cite{MLB}. 
    \item \textbf{Opaque Data Flow and Deferred Validation}. TE applications rely on data exchange between the host environment, the template engine, and the target consumer, which is often implicit and loosely typed~\cite{TypeJinja}. 
    As shown in Figure~\ref{fig: application-overview}, the host environment prepares a data context (e.g., Figure~\ref{fig: development-worflow} (b)) and passes it to the template engine, which then renders a template with this context to generate output for the target consumer to process. 
    Furthermore, the target consumer may trigger interactions with the host via requests. 
    Crucially, the validation of rendered results is typically deferred until it is processed by the target consumer (e.g., a browser). 
    This latency allows bugs originating from the host logic or template to remain latent until the execution of the target consumer. 
    The combination of implicit data flow and deferred validation makes it difficult to trace abnormal rendering result back to their root causes. 
    \item \textbf{Complex Integration}. Template engines are often deeply coupled with external frameworks. 
    This integration exacerbates the debugging complexity, since bugs may reside in the interplay between the engine and the framework, making them difficult to diagnose. 
    For instance, recent research on IaC ecosystems reveals that template-related issues are not only frequently discussed by practitioners~\cite{IaC2023, AnsibleChallenges} but also constitute a substantial portion of IaC program bugs~\cite{IaC2024}, highlighting the difficulty of managing template bugs within integrated environments. 
\end{itemize}

Despite the ubiquity of template engines and the complexity of their applications, bugs within TE applications remain under-investigated. 
Prior work primarily focuses on the security of template engines, especially the detection and defense of Server-Side Template Injection (SSTI) attacks~\cite{Kettle2015SSTI, 2024Survey, Zhao2023USENIX}. 
However, with the increasing adoption of template engines in mission-critical scenarios like high-traffic e-commerce websites and IT infrastructure management, TE application bugs deserve commensurate attention. 
While recent studies~\cite{CLB, MLB, 10.1002/smr.2507, Insight} have investigated cross-language bugs in Foreign Function Interface~\cite{FFI} (FFI, e.g., JNI~\cite{JNI} and CFFI~\cite{CFFI})-based multilingual software (FFI-MLS), TE applications represent a distinct class of multilingual software for three reasons: 
(1) Unlike the isolation of different languages in separate files in FFI-MLS, TE applications typically mix the templating language and target language within the same template. 
(2) In TE applications, the rendered results are further processed by the target consumer, whereas FFI-MLS directly executes cross-language function calls to obtain results. 
(3) Data is passed through implicit contexts in TE applications, which often leads to silent failures. 
In contrast, FFI-MLS rely on explicit parameters in FFIs, which are more likely to trigger immediate runtime exceptions upon failure. 
To summarize, while a comprehensive understanding of TE application bugs is urgently needed, such empirical evidence remains critically limited. 

To bridge this knowledge gap, we present an in-depth study to characterize bugs in TE applications, aiming to provide a systematic foundation for the research and development of specialized bug detection, automated repair, and static analysis techniques. 
Inspired by prior bug characterization studies (e.g., \cite{Deployment2021, DeepLearningCompiler, DataScience2025, DataVisualization2025, Container2024, GPU2025}), we propose the following research questions (RQs): 
\begin{itemize}[leftmargin=*, topsep=0pt]
    \item \textbf{RQ1 (Symptoms):} \textit{What are the symptoms of TE application bugs?} 
    Symptoms reveal how bugs manifest themselves during template rendering or consumer processing. 
    Categorizing these symptoms is essential for diagnosing and fixing TE application bugs in a more targeted way. 
    \item \textbf{RQ2 (Root Causes):} \textit{What are the root causes of TE application bugs?} 
    Root causes reflect the underlying issues leading to the observed symptoms. 
    Understanding these root causes and their relationship with symptoms is vital for designing effective diagnostic tools. 
    \item \textbf{RQ3 (Fix Patterns):} \textit{What fix patterns do developers employ to resolve TE application bugs?} 
    Fix patterns indicate what common approaches practitioners adopt to resolve bugs. 
    These insights can inform practitioners to avoid common pitfalls in TE application development and guide the development of automated repair and static analysis techniques. 
\end{itemize}

To answer these RQs, we systematically select 15 representative template engines across five programming languages. 
We curate a dataset comprising 1,004 application bugs related to these engines from Stack Overflow and follow the open coding procedure to identify their symptoms, root causes, and fix patterns. 
We further analyze the relationship between these dimensions. 
Our main findings include: 
(1) \textit{Abnormal Rendering Result} is the most frequent symptom, manifesting in nearly half (48.61\%) of the analyzed bugs. 
These bugs often manifest as silent failures by producing unexpected or blank output without an explicit crash, thereby significantly complicating bug localization. 
(2) \textit{Syntax Misuse} and \textit{Mismatched Data Context} are two major root causes, primarily stemming from misused template expressions or incorrect placeholder values in the data context. 
\textit{Incompatible Integration} also results in a substantial (16.73\%) of bugs. 
The root causes for \textit{Abnormal Rendering Result} are diverse, further hindering the diagnostic process. 
(3) The majority of fixes (67.92\%) are performed within the template, mostly through corrections to embedded logic or template syntax. 
However, a significant portion (20.67\%) of fixes occurs within the host environment, particularly to resolve \textit{Mismatched Data Context} bugs. 

Based on our findings, we propose actionable implications for tool designers, practitioners, and researchers. 
These include guidelines for robust template composition, essential features of IDE support tools, and promising directions for future research on automated detection, localization, and repair of TE application bugs. 
Furthermore, to demonstrate the practical utility of our findings, we develop two prototype tools for the Jinja engine: a syntax error detection and repair tool, and a template element extractor designed to facilitate data tracking and control-flow analysis. 
We intend to integrate them into a VS Code extension to facilitate TE application development in the future. 

In summary, this paper makes the following contributions:
\begin{itemize}[leftmargin=*, topsep=0pt]
    \item To the best of our knowledge, we conduct the first large-scale study to characterize bugs in TE applications, based on 1,004 bugs across 15 template engines. 
    \item We present three taxonomies to comprehensively characterize the symptoms, root causes, and fix patterns of TE application bugs. We also reveal their correlations. 
    \item We distill a series of implications for the composition of templates, the design of future IDE support, and future research on specialized automated debugging techniques. Furthermore, we implement two prototype tools for the Jinja template engine. 
    \item We have made our curated dataset and prototype tools publicly available at ~\cite{template_engine_bugs_2026} to facilitate reproducibility and future research.
\end{itemize}

\section{Background}\label{s: template_engine}
A template engine is a specialized software library implemented in a specific programming language (e.g., Python for Jinja, Java for Thymeleaf). 
It provides: 1) a domain-specific templating language for composing templates, and 2) Application Programming Interfaces (APIs) for environment configuration and data injection. 
Therefore, it can only be used in the host environment developed with the same language. 
In this paper, we refer to the process of leveraging template engines to generate dynamic documents as \textit{TE application development}. 

\textbf{Distinction from Scaffolding~\cite{Scaffold} and Code Snippets~\cite{Snippet}.} 
While the term ``template'' is used broadly in software engineering, template engines are fundamentally distinct from development-time utilities such as project scaffolding and code snippets in three key dimensions: 
(1) \textit{Operation Phase}. Template engines operate during runtime (i.e., execution phase) to generate documents. 
In contrast, scaffolding is utilized during the initialization phase to bootstrap a project structure setup, and snippets are used during the coding phase as interactive aids for repetitive code structure. 
(2) \textit{Data-driven Generation}. Template engines automatically and repeatedly generate documents based on shifting data contexts provided by the host environment. 
Scaffolding and snippets are typically one-off static operations, where the generated or inserted content require further human modification before they are functional. 
(3) \textit{Output Type}. The output of a template engine is typically a complete and structured document (e.g., an HTML page or a YAML config) intended for immediate consumption by a downstream system. 
Conversely, scaffolding produces a project-level directory skeleton and snippets produce granular source code fragments. 
These differences underscore the inherent complexities of TE application bugs. 

Figure~\ref{fig: development-worflow} illustrates the typical workflow of TE application development, which consists of five stages: 
\begin{enumerate}[leftmargin=*, topsep=0pt]
    \item \textit{Template Engine Configuration}. This stage involves initializing the engine, resolving its environment-specific dependencies, and usually configuring its integration with external frameworks (e.g., Flask or Spring Boot) used in the host environment. 
    \item \textit{Template Composition}. In this stage, developers compose templates using the templating language provided by the engine. 
    A template is a polyglot artifact characterized by syntactic interleaving, where dynamic logic (directives written in the templating language) is woven into static boilerplate (target-language content, e.g., HTML). 
    As shown in the Jinja template example in Figure~\ref{fig: development-worflow} (a), a templating language typically supports four core primitives: 
    \begin{itemize}[leftmargin=*, topsep=0pt]
        \item \textbf{Delimiters} are markers (e.g., \Code{\{\{ \}\}}) that separate dynamic logic from static boilerplate. 
        \item \textbf{Placeholders} are variables (e.g., \Code{username}) that act as receptors for data injected from the host environment. 
        \item \textbf{Expressions} are operations used to transform or format data (e.g., \Code{todo.finished == 100}). 
        \item \textbf{Control Structures} are directives (e.g., \Code{for} and \Code{if}) that implement control logic, such as conditionals, loops, filters, and template inheritance. 
        They distinguish template engines from string interpolation features in general-purpose languages (e.g., Python f-strings~\cite{FStrings}) by enabling complex and flexible logic within the template. 
    \end{itemize}
    \item \textit{Template Compilation}. Modern template engines are generation-based for performance~\cite{Zhao2023USENIX}. 
    In this stage, the engine parses the template and generates an Intermediate Representation (IR). 
    For example, Jinja compiles templates into Python code~\cite{JinjaIR}. 
    This IR is typically cached to minimize overhead in subsequent repetitive rendering operations. 
    \item \textit{Template Rendering}. In this stage, the engine receives a \textit{data context} from the host environment (e.g., Figure~\ref{fig: development-worflow} (b)), which specifies concrete values for placeholders in the template. 
    It then binds this data to placeholders in the IR, executes the embedded logic, and produces the final output (e.g., the HTML document in Figure~\ref{fig: development-worflow} (c)). 
    \item \textit{Rendered Result Processing}. In the final stage, the target consumer (e.g., a browser or SQL engine) processes the rendered document. 
    Since the engine is agnostic to the target language semantics, it may successfully render a document that is syntactically invalid or semantically incorrect for the consumer (e.g., broken HTML tags). This leads to deferred validation, where defects remain latent until the execution phase within the consumer.
\end{enumerate}

\section{Methodology}\label{s: methodology}
This section details our methodology, which consists of three steps: (i) selection of representative template engines (Section~\ref{ss: engine_selection}), (ii) collection of TE application bugs (Section~\ref{ss: bug_collection}), and (iii) manual labeling to identify their symptoms, root causes, and fix patterns (Section~\ref{ss: manual_labeling}). 

\subsection{Template Engine Selection}\label{ss: engine_selection}
As noted in Section~\ref{s: template_engine}, template engines are implemented as software libraries. 
Given the vast array of available template engines~\cite{2024Awesome, 2024Survey}, it is infeasible to analyze all of them. 
To ensure our study covers the most prevalent and representative engines used in practice, we first identified a candidate set of template engines from the community-curated collection~\cite{2024Awesome} and prior work~\cite{2024Survey}. 
We then quantified their practical adoption and community discussion via Stack Overflow. 
We did not use GitHub-based metrics, such as the number of stars or dependent repositories~\cite{githubdependencygraph}, since many prominent engines do not have independent repositories. 
For instance, Django-Template is part of the Django repository, and Blade is part of the Laravel repository. 
To ensure sufficient bugs for our analysis, we selected only those engines with at least 1,000 questions as of June 2025. 
This process yields 15 template engines, spanning five major programming languages, as detailed in Table~\ref{tab: dataset_statistics}. 
The templating languages provided by these engines all support the four primitives listed in Section~\ref{s: template_engine} but with different syntax symbols. 

\begin{table}[ht]
\centering
\caption{The statistics of selected template engines. The Language column denotes the implementation language of the engine, which, for all studied engines, is identical to the language used in the host environment. }
\label{tab: dataset_statistics}
\renewcommand{\arraystretch}{1.1}
\setlength{\tabcolsep}{2pt}
\begin{minipage}{0.49\textwidth}
    \begin{tabular}{lrrr}
    \toprule
    \textbf{Template Engine} & \textbf{Language\tnote{*}} & \textbf{\# Questions} & \textbf{\# Bugs} \\
    \midrule
    Django-Template~\cite{Django-Template} & Python & 19,627 & 327 \\
    Jinja~\cite{Jinja} & Python & 9,747 & 178 \\
    Thymeleaf~\cite{Thymeleaf} & Java & 9,295 & 143 \\
    FreeMarker~\cite{FreeMarker} & Java & 3,107 & 14 \\
    Velocity~\cite{Velocity} & Java & 2,373 & 7 \\
    Blade~\cite{Blade} & PHP & 7,470 & 72 \\
    Twig~\cite{Twig} & PHP & 10,619 & 46 \\
    Smarty~\cite{Smarty} & PHP & 4,477 & 5 \\
    \bottomrule
    \end{tabular}
\end{minipage}
\hfill
\begin{minipage}{0.49\textwidth}
    \begin{tabular}{lrrr}
    \toprule
    \textbf{Template Engine} & \textbf{Language\tnote{*}} & \textbf{\# Questions} & \textbf{\# Bugs} \\
    \midrule
    Handlebars.js~\cite{Handlebars.js} & JavaScript & 7,690 & 33 \\
    Pug~\cite{Pug} & JavaScript & 6,367 & 33 \\
    Mustache.js~\cite{mustache.js} & JavaScript & 1,958 & 6 \\
    EJS~\cite{EJS} & JavaScript & 6,475  & 97 \\
    Liquid~\cite{Liquid} & Ruby & 4,045  & 32 \\
    Haml~\cite{Haml} & Ruby & 3,540 & 2 \\
    ERB~\cite{ERB} & Ruby & 2,632 & 9 \\
     & & & \\
    \bottomrule
    \end{tabular}
\end{minipage}
\end{table}

\subsection{Bug Collection}\label{ss: bug_collection}
Following established empirical practices~\cite{Tensorflow2018, DeepLearing2019, DLSPB, DLStack, RLBugs}, we chose Stack Overflow (SO) as our primary data source. 
As a premier Q\&A platform, SO has gathered over 24 million questions arising from practical software development scenarios~\cite{7551996} and over 31 million users with diverse professional backgrounds, from novices to industry experts and open-source contributors. 
Prior work has revealed that practitioners frequently utilize SO to seek solutions for fixing bugs~\cite{Treude11ICSE, Xia2017EMSE}. 
Furthermore, SO employs multiple quality-control mechanisms, such as community voting and reputation systems, and utilizes a structured tagging system to facilitate the identification of relevant technical issues. 
Therefore, curating bugs from SO is an established strategy for characterizing bugs in real-world applications within specific domains (e.g., \cite{DLSPB, DLStack, DeepLearing2019}). 

We downloaded the Stack Overflow data dump released on June 30, 2025, which records comprehensive metadata for each question, including tags, title, body, and score. 
Notably, each SO question has one to five tags indicating related concepts and technologies. 
For each of the 15 selected template engines, we identified its official or most relevant tag (typically the engine’s name) and iterated through the data dump to extract all associated questions based on whether they contain the identified tag. 
To ensure our dataset reflects recent development practices and remains relevant to recent engine versions, we retained only questions created after January 1, 2020~\cite{DLSPB, DataVisualization2025}. 

Recognizing that SO questions can vary in complexity, we applied a multi-stage filtering process to ensure our dataset captures non-trivial, high-quality bugs rather than simple ``how-to'' questions~\cite{DeepLearing2019, Deployment2021, DLSPB, DLStack, Docker2025}. 
Specifically, we excluded questions that: (i) lacked an \textit{accepted answer}, (ii) did not contain source code snippets, or (iii) had a non-positive score. 
To further isolate genuine bugs, we performed keyword-based filtering, retaining only questions that contain bug-related words (e.g., ``bug'', ``not working''). 
This process yielded an initial set of 2,501 questions. 

The first two authors then conducted a manual review to identify genuine TE application bugs. 
In an initial pilot round, they randomly sampled 50 questions for independent labeling. 
The inter-rater agreement measured by Cohen’s Kappa coefficient ($\kappa$)~\cite{Cohen} was 0.52, indicating moderate agreement. 
The authors then discussed the conflicts to unify the labeling criteria, for instance, excluding questions related to general business logic errors or with ambiguous symptom descriptions. 
Following the refined criteria, a second round labeling of another 50 sampled questions yielded a $\kappa$ score of 0.88, suggesting an almost perfect agreement. 
Subsequently, the two authors independently labeled the remaining questions, achieving a $\kappa$ of 0.83. 
The disagreements were resolved with another discussion. 

Ultimately, we obtained a dataset consisting of 1,004 TE application bugs, whose scale is comparable to prior studies~\cite{DLSytemFaults, JupyterNotebbokBug}. 
The number of collected bugs for each template engine is shown in Table~\ref{tab: dataset_statistics}. 
The dataset exhibits significant technical depth and complexity, with the mean and median lines of code (LOC) of 38 and 23 respectively, substantially higher than those of general Stack Overflow questions (23 and 9). 
Furthermore, the dataset offers rich contexts: 81.5\% of the bugs contain multiple code snippets (e.g., a template code snippet and a host code snippet), and 17.9\% include visual screenshots of the rendered result. 
Although this dataset may include some relatively simple bugs, we intentionally retained them as they represent the diverse real-world hurdles faced by practitioners and highlight critical usability gaps in current development support tools~\cite{Tensorflow2018}. 

While SO serves as our primary data source, we recognize the potential platform bias. 
Therefore, we further collected bugs from GitHub repositories and validated the taxonomies derived from SO against them, as detailed in Section~\ref{ss: taxonomy_validation}. 

\subsection{Manual Labeling}\label{ss: manual_labeling}
To address our research questions, we performed a systematic manual labeling of the symptoms, root causes, and fix patterns for all 1,004 bugs. 
Before labeling, the first two authors (i.e., labelers) familiarized themselves with each of the 15 template engines by scrutinizing their official documentation, mastering their templating language syntax, and experimenting with representative code examples. 

Following established empirical methodologies~\cite{Deployment2021, SoftwareRelease2022, DLStack}, we conducted a pilot labeling to derive initial taxonomies. 
We selected all bugs with a score $\geq 2$ across all the 15 engines, resulting in a pilot set of 338 bugs. 
We then followed an open coding procedure~\cite{OpenCoding} to analyze these bugs. 
The two labelers independently performed an exhaustive review of each bug post, including the title, question description, comments, and answers. 
During this process, they annotated text segments describing observable symptoms (primarily from questions and associated comments) and root causes or fixes (primarily from answers and associated comments), assigning initial low-level codes to each dimension. 
Subsequently, the labelers took an iterative, bottom-up inductive process to construct the taxonomies. 
They first reconciled their initial codes, unifying synonymous terms and deliberating on any discrepancies until a consensus was reached. 
They then employed the constant comparison method, i.e., taking one bug at a time and comparing its code with existing categories, to determine whether codes should be merged or maintained as distinct categories. 
Throughout this process, an external arbitrator with extensive experience in TE application development was involved to resolve conflicts. 

To ensure the reproducibility and verifiability of our results, the two labelers collaboratively developed a comprehensive coding guide, which was further reviewed by the arbitrator. 
Using this guide, the two labelers independently re-labeled the 338 pilot bugs to verify consistency, achieving $\kappa$ scores of 0.95, 0.87, and 0.84 for symptoms, root causes, and fix patterns, respectively. 
The disagreements were resolved by involving the arbitrator to further refine the guide. 
Finally, the two labelers independently labeled the remaining 666 bugs, maintaining high $\kappa$ scores of 0.94 (symptoms), 0.83 (root causes), and 0.82 (fix patterns) and resolving the disagreements. 
The high inter-rater reliability underscores the stability of our taxonomies and the reproducibility of our results. 

\section{Results}\label{s: results}
\subsection{RQ1: Symptoms}\label{ss: rq1-results}
\begin{figure}
    \centering
    \includegraphics[width=\linewidth]{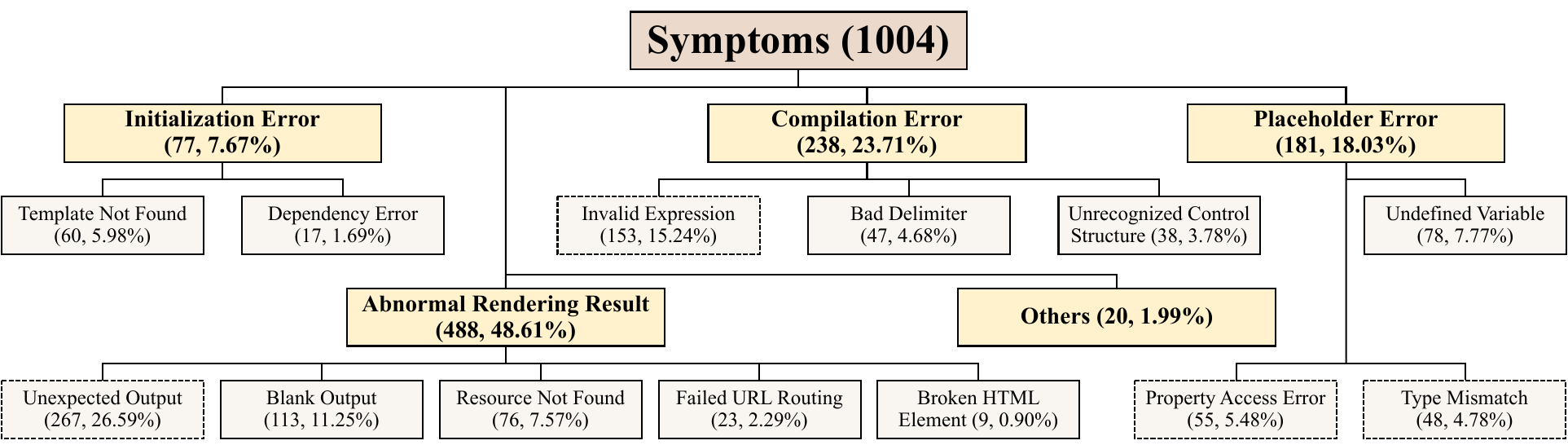}
    \caption{The taxonomy and distribution of bug symptoms in TE applications. Dotted rectangles and solid rectangles represent symptoms shared with cross-language bugs and those unique to TE applications, respectively.}
    \label{fig: symptoms}
\end{figure}

Figure~\ref{fig: symptoms} illustrates the taxonomy and distribution of bug symptoms in TE applications, which is structured into five high-level categories and 13 leaf symptoms. 

\textbf{Initialization Error.} To ensure the proper operation of a template engine, the host environment needs to correctly specify the template file location and install required dependencies. 
77 (7.67\%) of the analyzed bugs manifest as runtime exceptions during this initialization phase. 
Specifically, 60 (5.98\%) bugs show \textit{Template Not Found} errors, where the engine fails to resolve the path to the template file. 
These bugs occur predominantly when the engine is integrated with external frameworks (e.g., Django and Spring Boot) that impose specific directory structures. 
Besides, several (17, 1.69\%) bugs exhibit \textit{Dependency Error}, where the host environment fails to import the template engine or necessary plugins. 

\textbf{Compilation Error.} Templates incorporate logic expressed through domain-specific templating languages. 
238 (23.71\%) of the analyzed bugs manifest as syntax errors when the engine compiles the template into an Intermediate Representation (IR), making it the second most common symptom. 
The significant prevalence of compilation errors suggests that practitioners face substantial hurdles with fundamental templating language syntax. 
Furthermore, this finding reveals a critical gap in current IDE support, specifically the lack of informative diagnostic feedback and automated fix mechanisms tailored for templating languages. 
As introduced in Section~\ref{s: template_engine}, the compilation stage primarily validates three primitives: delimiters, expressions, and control structures. 
Specifically, 153 (15.24\%) of the bugs exhibit \textit{Invalid Expression} errors, where the engine fails to parse specific tokens within the expression, such as nested interpolations, arithmetic operations, or conditional constructs. 
Additionally, 47 (4.68\%) of the bugs manifest as \textit{Bad Delimiter} errors where templates contain unclosed or mismatched delimiters. 
For example, a bug~\id{EJS}{60225400}\footnote{In the remainder of this paper, we use (Engine, PostID) to refer to a bug in our dataset, denoting that the bug occurred in the application of <Engine> and corresponds to the SO question at \url{https://stackoverflow.com/questions/<PostID>}.} in our dataset shows the error message: \textit{``Could not find matching close tag for <\%''}. 
Finally, 38 (3.78\%) of the bugs involve \textit{Unrecognized Control Structure} errors, where the engine fails to recognize specific tags or filters. 
These bugs typically arise when developers attempt to use customized extensions without adhering to the engine's formal registration protocols. 
Representative examples include failures to resolve unregistered tags (e.g., \textit{mes\_tags is not a registered tag library''}~\id{Django-Template}{59932340}) or unknown filters (e.g., \textit{Unknown "format\_number" filter on Symfony 4 project''}~\id{Twig}{63018096}).

\textbf{Placeholder Error.} 
Following the compilation stage, the engine binds placeholders to concrete values retrieved from the data context and executes the template logic to render the final output. 
We identify 181 (18.03\%) bugs that throw runtime exceptions during the resolution and manipulation of these placeholders, manifesting as three distinct symptoms. 
Specifically, \textit{Undefined Variable} is the most prevalent symptom in this category, covering 78 (7.77\%) bugs. 
Exceptions are thrown when the engine is unable to resolve a value for a specific placeholder from the data context. 
Even when a value is successfully resolved, subsequent operations may fail if the template attempts to perform invalid property access on the placeholder object, i.e., \textit{Property Access Error} manifested in 55 (5.48\%) bugs. 
These exceptions frequently involve dereferencing properties on null objects, e.g., \textit{``Property or field 'author' cannot be found on null''}~\id{Thymeleaf}{64224649}, or attempting to access restricted properties, e.g., \textit{``Access has been denied to resolve the property "from"''}~\id{Handlebars.js}{59690923}. 
Finally, exceptions also arise in 48 (4.78\%) bugs when the data type of the resolved placeholder value is incompatible with the requirements of a template operation, i.e., \textit{Type Mismatch}. 
Examples include attempting to iterate over a non-iterable object, e.g., \textit{``datetime.date object is not iterable''}~\id{Django-Template}{77934119}, or performing arithmetic with incompatible operands, e.g., \textit{``unsupported operand type(s) for /: 'decimal.Decimal' and 'float'''}~\id{Jinja}{73315956}.  

\textbf{Abnormal Rendering Result.} 
Even if the engine renders the template without runtime exceptions, the generated output may diverge from the developer's expectations once processed by the target consumer. 
We categorize these bugs as \textit{Abnormal Rendering Result}. 
This is the most prevalent symptom in our dataset, accounting for 488 (48.61\%) of the bugs. 
Due to the deferred validation nature of TE applications, these silent bugs often bypass the \textit{Template Rendering} stage undetected and only manifest in the \textit{Rendered Result Processing} stage within the consumer. 
This category is further divided into five leaf symptoms:
\begin{itemize}[leftmargin=*, topsep=0pt]
    \item \textit{Unexpected Output (267, 26.59\%)}: This is the most frequent symptom overall, where the engine successfully evaluates a template expression, yet the result diverges from the intended value. For example, a developer reported that an expression produced \textit{`foo'} when the expected output was \textit{`bar'}~\id{Liquid}{68623329}. 
    \item \textit{Blank Output (113, 11.25\%)}: In these cases, the engine fails to evaluate an expression, resulting in an empty output, e.g., \textit{``{{ partner.title }} does not grab the Post Title, it's blank''}~\id{Twig}{61805121}. 
    \item \textit{Resource Not Found (76, 7.57\%)}: The most common scenario where template engines are used is to produce HTML documents. 
    These bugs occur when the rendered HTML documents fail to load external assets in the browser, such as CSS, JavaScript, or images. 
    While some of these errors manifest in the browser’s console, e.g., \textit{``GET .../images/normal.gif 404 (Not Found)''}~\id{Handlebars.js}{71242969}, others remain latent until the user observes missing content, e.g., \textit{``the user.jpg image is not displayed''}~\id{Pug}{79520210}. 
    \item \textit{Failed URL Routing (23, 2.29\%)}: 
    When integrated with web frameworks, template engines utilize dynamic abstractions (e.g., \Code{\{\% url \%\}} in Django-Template) rather than hardcoded paths to resolve links via logical names. 
    This mechanism allows the host environment to manage URL patterns independently of the presentation layer (i.e., templates). 
    However, it establishes an implicit contract between templates and the framework's routing table: any mismatch in logical names or required parameters triggers \textit{Failed URL Routing} bugs, which typically manifest as HTTP errors (e.g., 404) or framework-specific runtime exceptions (e.g., \textit{NoReverseMatch} in Django). 
    Unlike \textit{Template Not Found} bugs which are engine-level exceptions thrown during initialization, these bugs exemplify deferred validation, as they are only triggered during the rendered result processing stage when the host attempts to resolve the dynamic link. 
    \item \textit{Broken HTML Element (9, 0.90\%)}: A small subset of bugs manifest as visual or functional deficiencies in specific HTML elements, such as the failure to render specific elements or the loss of interactivity. 
\end{itemize} 

\textbf{Comparison with Cross-Language Bug Symptoms.} 
To highlight the distinct nature of TE application bugs, we compare our symptom taxonomy with that of cross-language bugs (CLBs) established in prior work~\cite{CLB}. 
In Figure~\ref{fig: symptoms}, we differentiate between shared symptoms and those unique to TE applications with dotted and solid borders, respectively. 
While our taxonomy shares four leaf symptoms with~\cite{CLB}, the remaining nine symptoms are unique to TE applications and account for 481 (47.91\%) of the analyzed bugs. 
This significant difference stems from the architectural distinctions between FFI-based multilingual software and TE applications discussed in Section~\ref{s: intro}, such as polyglot templates, implicit data exchange, deferred validation, and complex integration. 
These characteristics introduce unique symptoms, such as \textit{Bad Delimiter}, \textit{Undefined Variable}, \textit{Resource Not Found}, and \textit{Initialization Error}. 

\begin{figure}
    \centering
    \includegraphics[width=0.6\linewidth]{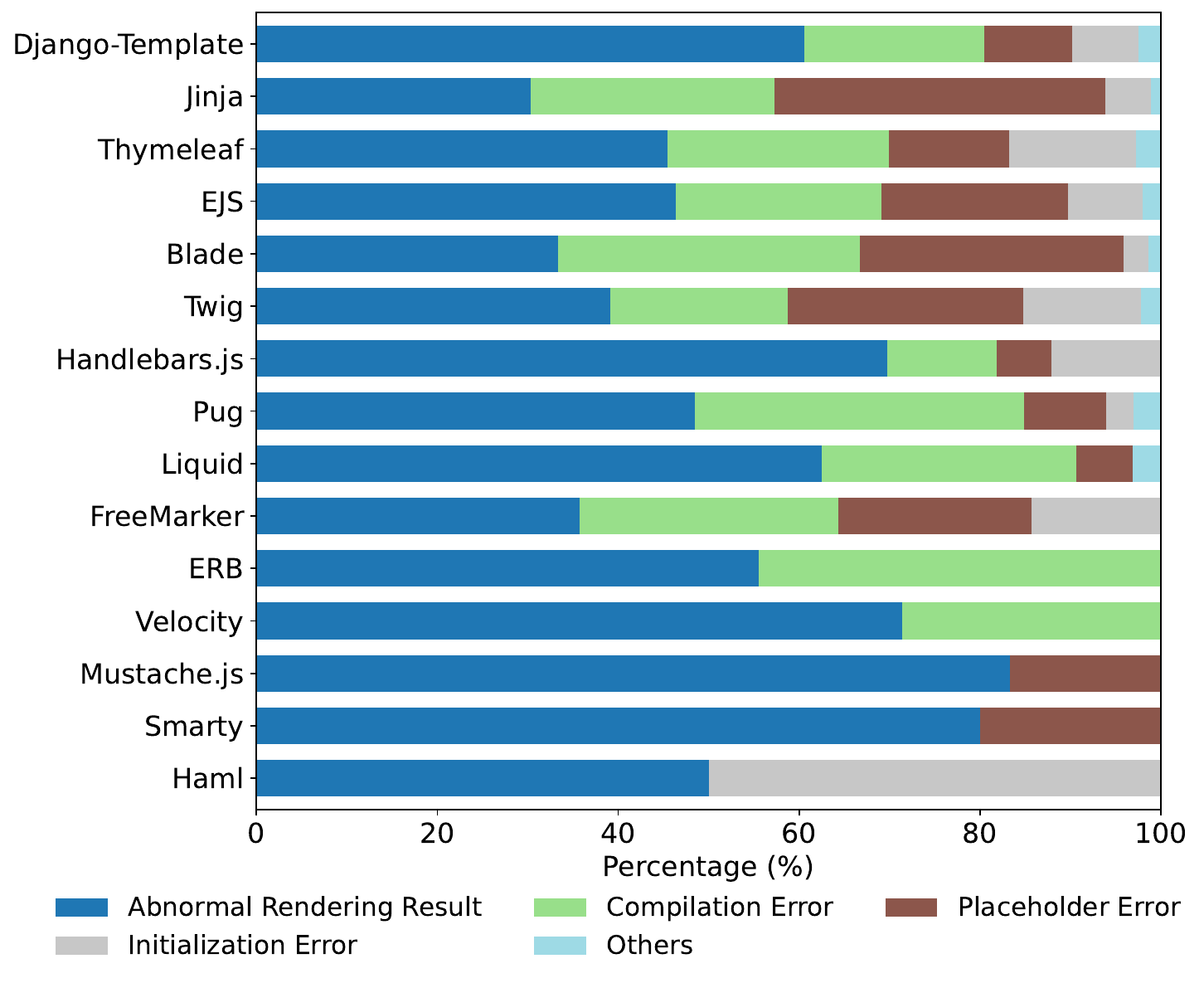}
    \caption{The symptom distribution of TE application bugs by engine.}
    \label{fig: te-symptom}
\end{figure}

\textbf{Distribution of Symptoms across Template Engines.} 
We further analyzed the symptom distribution of TE application bugs across the 15 studied template engines. 
As shown in Figure~\ref{fig: te-symptom}, we observe that \textit{Abnormal Rendering Result}, \textit{Compilation Error}, and \textit{Placeholder Error} are the three dominant symptoms for all subjects, collectively accounting for over 80\% of the bugs. 
Specifically, \textit{Abnormal Rendering Result} is the most prevalent symptom for 14 of the 15 engines. 
A notable exception is Jinja, where \textit{Placeholder Error} (36.52\%) slightly exceeds \textit{Abnormal Rendering Result} (30.34\%). 
This shift likely stems from Jinja's widespread adoption in non-web domains, such as Infrastructure-as-Code and data engineering, where complex, multi-layered data contexts make placeholder resolution more error-prone than in standard web views. 
Furthermore, we find that engines tightly coupled with web frameworks (e.g., Django-Template/Django, Thymeleaf/Spring Boot, and Twig/Symfony) exhibit a higher frequency of \textit{Initialization Error} bugs. 
This may be attributed to the ``convention over configuration''~\cite{CoC} design paradigm of these frameworks, where developers often struggle with implicit template loading and path resolution rules. 

Despite these nuances, a Kruskal-Wallis H-test~\cite{Kruskal01121952} confirms that the symptom distribution across the 15 engines exhibits no statistically significant variance ($p > 0.05$). 
This demonstrates that our taxonomy captures the fundamental, intrinsic bug characteristics of TE applications rather than being biased toward a specific engine or ecosystem.

\begin{findingbox}
\textbf{Summary for RQ1:} Bugs in TE applications mainly manifest 13 symptoms across four categories. 
\textit{Abnormal Rendering Result} is the most common category (48.61\%) which often manifests as silent failures such as unexpected or blank output. 
\textit{Compilation Error} are the second most common category (23.7\%) followed by \textit{Placeholder Error} (18.03\%) and \textit{Initialization Error} (7.67\%). 
The symptom distribution across the studied template engines does not show statistically significant difference. 
\end{findingbox}

\subsection{RQ2: Root Causes}\label{ss: rq2-results}
\begin{figure}
    \centering
    \includegraphics[width=\linewidth]{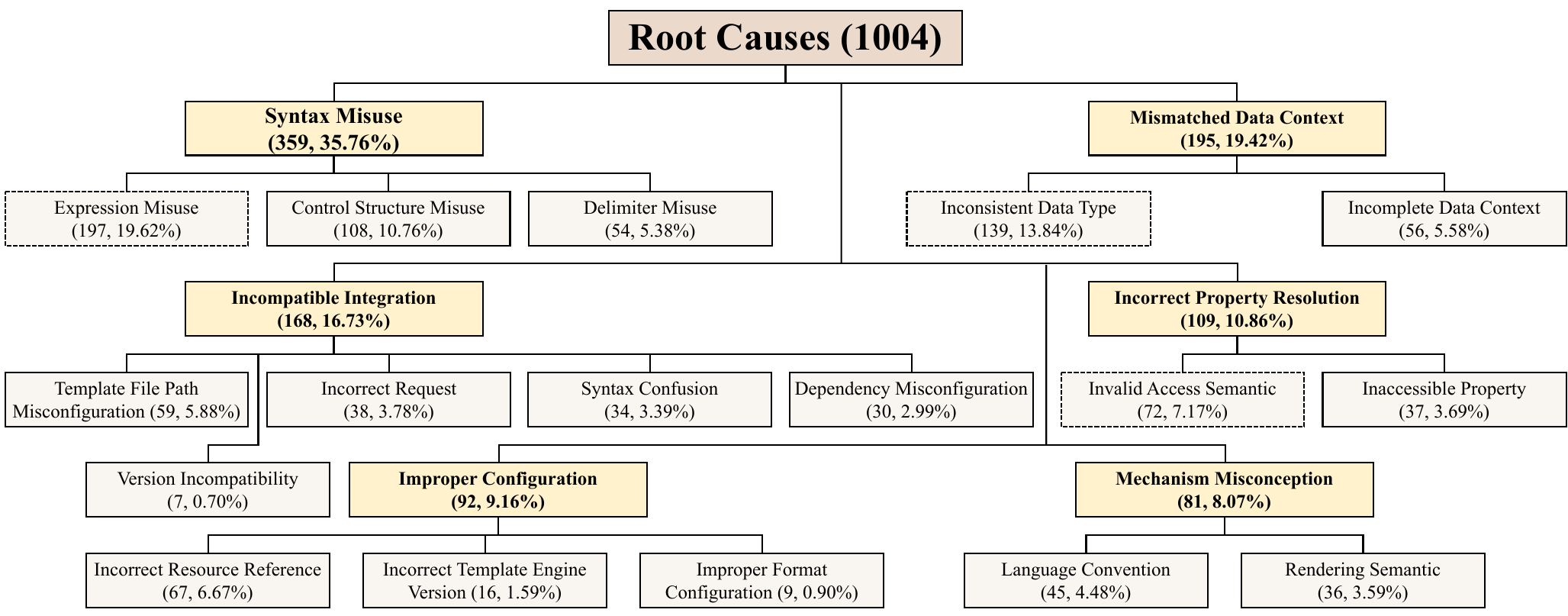}
    \caption{The taxonomy and distribution of bug root causes in TE applications. Dotted rectangles and solid rectangles represent root causes shared with CLBs and those unique to TE applications, respectively.}
    \label{fig: root-causes}
\end{figure}

Figure~\ref{fig: root-causes} illustrates the taxonomy and distribution of bug root causes in TE applications, which is structured into six high-level categories and 17 leaf root causes. 

\textbf{Syntax Misuse.} 
This category consists of bugs arising from various incorrect usage of templating language syntax, ranging from grammar violations to the misuse of similar syntax. 
It is the most common root cause in our dataset, accounting for 358 (35.66\%) of the bugs. 
While each engine provides a domain-specific templating language, our findings suggest that developers frequently struggle with the inflexibility and syntax nuances of these languages. 
We further divide this category into three leaf categories based on the primitives. 

\textit{Expression Misuse} refers to grammatical violations of the templating language. 
It occurs when developers attempt to use operators, function calls, or logic structures that are not supported by the engine's formal grammar. 
It is the most common root cause, accounting for 197 (19.62\%) of the analyzed bugs. 
These bugs are typically caught by the engine's parser, resulting in various compilation errors. 
Unlike general-purpose languages, templating languages often impose strict syntax constraints. 
For example, although developers can pass Python objects to a Django template, they cannot manipulate them as freely as they can in Python code. Specifically, a bug~\id{Django-Template}{71449049} was caused because \textit{``Django's template language does not allow function calls, subscripting, operators, etc''}. 
Another bug~\id{Thymeleaf}{69276002} was caused by the constraint that \textit{``Thymeleaf does not support the use of expressions for evaluating URL fragments''}. 
Furthermore, syntactic inconsistencies across data types can be non-intuitive. 
For instance, in Jinja, while the \Code{+} operator is used to concatenate numbers or lists, the tilde ($\sim$) operator is the preferred syntax for string concatenation, which results in the bug~\id{Jinja}{72176681}. 

\textit{Control Structure Misuse} comprises bugs occurred when developers violate the usage specification of tags or filters provided by the engine and accounts for 108 (10.76\%) of the analyzed bugs. 
In some cases, developers place tags in the wrong place. 
For instance, the \Code{extends} tag must be the first tag in a Jinja template, and thus \textit{``put the extends template tag in an if-else''} causes a bug~\id{Jinja}{67490509}. 
In other cases, developers provide an incorrect number or type of arguments to template filters, e.g., the \Code{\{math\}} tag in Smarty requires all arguments to be pure numeric types, otherwise, an exception was thrown~\id{Smarty}{60286749}. 

\textit{Delimiter Misuse} involves the incorrect selection or usage of delimiters and accounts for 54 (5.38\%) of the analyzed bugs. 
Delimiters are symbols that separate dynamic logic from boilerplate content. 
Template engines usually provide multiple delimiters for different purposes, such as printing output and executing logic. 
For instance, the \Code{<\% \%>} delimiter in ERB is for executing the bracketed Ruby statement, while the \Code{<\%= \%>} delimiter is for evaluating the bracketed Ruby expression and printing the result. 
Therefore, \Code{<\%= if flash[:notice] \%>} raises an exception in bug~\id{ERB}{63495203}. 
Additionally, many engines prohibit the \textbf{nesting of delimiters} (e.g., \Code{\{\{ foo['\{\{ key \}\}'] \}\}}), a syntactic constraint that developers frequently overlook across ecosystems. 
This constraint is explicitly cited as a root cause in numerous bugs. 
For instance, developers were cautioned that \textit{``You can't nest <\% \%> blocks''}~\id{EJS}{60225400}, 
\textit{``Handlebars does not support nesting mustache expressions''}~\id{Handlebars.js}{67129074}, 
and \textit{``mustache do not stack''}~\id{Jinja}{67408680}. 

\textbf{Mismatched Data Context.} 
Since template files are usually separated from the host environment, the data context prepared by the host frequently mismatches the semantic expectations of the template placeholders. 
This category exemplifies the \textit{Opaque Data Flow} characteristic of TE applications discussed in Section~\ref{s: intro}, where the lack of an explicit data contract between the host and the template hinders the correct preparation of data contexts. 
It is the second most common root cause in our dataset, accounting for 195 (19.42\%) of the bugs and comprising two leaf categories. 
\textit{Inconsistent Data Type} is the second most common root cause with 139 (13.84\%) of the analyzed bugs, following \textit{Expression Misuse}. 
Unlike syntax errors, these bugs manifest in the template rendering stage. 
In these cases, the host environment injects a value into the data context that is of an incompatible type for the intended template operation. 
For instance, in bug~\id{Django-Template}{73077282}, the host-side logic mistakenly passed a \Code{QuerySet} object (a collection of database records) to a placeholder which expects a single record instead. 
Here, the template logic is correct, but it cannot execute because the injected data deviates from the expected schema. 
We found that 56 (5.58\%) of the bugs stem from an \textit{Incomplete Data Context}, where the host environment fails to provide values for required placeholders. 
This omission precludes the engine from resolving these placeholders during the rendering stage, typically manifesting as the \textit{Undefined Variable} symptom. 

\textbf{Incompatible Integration.} 
168 (16.73\%) of the analyzed bugs stem from conflicting or erroneous integrations between the template engine, the host environment, and the target consumer. 
It echoes the \textit{Complex Integration} nature of TE applications discussed in Section~\ref{s: intro}, where engines are often integrated with web frameworks in practice. 
This category comprises five leaf categories. 
\begin{itemize}[leftmargin=*, topsep=0pt]
    \item \textit{Template File Path Misconfiguration.} To successfully resolve template file paths, developers must either adhere to the framework's default discovery paths or explicitly configure custom template directories. 
    Otherwise, a \textit{Template Not Found} exception is raised. 
    For example, Spring Boot looks for Thymeleaf templates in \Code{src/main/resources/templates} by default, while Django automatically searches directories mapped to registered applications. 
    Notably, nearly all (59 of 60) bugs with the symptom of \textit{Template Not Found} originated from this root cause. 
    \item \textit{Incorrect Request.} As noted in Section~\ref{s: intro}, target consumers often interact with the host by issuing requests parameterized via placeholders within the template. 
    Successful data exchange across this boundary requires strict naming alignment between template-side placeholders and the corresponding host-side function arguments or object properties. 
    Violations of this implicit naming contract result in 38 (3.78\%) of the analyzed bugs. 
    For instance, a bug~\id{Django-Template}{66411846} occurred because a parameter was named \Code{post\_id} in the template but expected as \Code{video\_id} by the host-side controller. 
    \item \textit{Syntax Confusion.} 34 (3.39\%) of the analyzed bugs occur when developers conflate the syntax rules of the host or target language with those of the templating language. 
    Examples include using HTML-style comments (\Code{<!-- -->}) instead of engine-specific comment tags, which results in unintended template logic execution~\id{Django-Template}{69459997} or attempting to use Python modules directly within a Jinja template expression~\id{Jinja}{63118911}. 
    This category further highlights the Multi-Language Composition complexity of TE applications discussed in Section~\ref{s: intro}, where developers must navigate overlapping syntactic boundaries. 
    \item \textit{Dependency Misconfiguration.} 30 (2.99\%) of the bugs arise from misconfigured environment dependencies, most notably security permissions and service mappings within the host framework. 
    For example, restrictive Spring Security configurations can inadvertently block access to template-referenced assets like CSS directories~\id{Thymeleaf}{59695497}. 
    \item \textit{Version Incompatibility.} 7 (0.70\%) bugs are attributed to version conflicts between the engine and its associated plugins or the framework. 
\end{itemize}

\textbf{Incorrect Property Resolution.} 
The host environment frequently injects complex objects (e.g., Java beans, dictionaries, or database models) to the template. 
We identified 109 (10.86\%) bugs stemming from erroneous attempts to dereference properties within these objects. 
Crucially, unlike the \textit{Mismatched Data Context} category which refers to the host providing invalid data, this category focuses on cases where the host environment provides the correct data objects, yet the template logic fails to correctly resolve them. 
This category comprises two leaf root causes. 
\textit{Invalid Access Semantic} involves 72 (7.17\%) bugs where the template employs an incorrect approach to dereference property values from host-injected objects, e.g., incorrectly using placeholders to dynamically construct property names~\id{Twig}{63140944}. 
In contrast to \textit{Expression Misuse} bugs which violate the engine's formal grammar and trigger immediate compilation errors, this root cause involves expressions that are syntactically well-formed but semantically incompatible with the object's resolution rules. 
Consequently, these bugs typically remain latent until the template rendering stage. 
\textit{Inaccessible Property} occurs when the template attempts to access properties that either do not exist in the object or are restricted due to visibility constraints, such as private properties~\id{Blade}{66629537}. 
It accounts for 37 (3.69\%) bugs and further highlights the opaque data flow nature of TE applications. 

\textbf{Improper Configuration.} 
Template engines require proper configurations to work as intended. 
This category involves 92 (9.16\%) bugs stemming from erroneous or missing template engine configurations and comprises three leaf root causes. 
Over two-thirds (67, 6.67\%) of bugs in this category fall into \textit{Incorrect Resource Reference}, which involves the misconfiguration of static asset (e.g., CSS files or images) folders or the application of improper path referencing strategies. 
Notably, many engines mandate a dedicated asset root and requires assets to be referenced via absolute paths relative to the root rather than the template's location. 
Email templates represent a significant edge case where both relative and absolute paths typically fail. 
In such scenarios, assets must be either served via remote URLs or embedded as attachments~\id{Handlebars.js}{76755262}. 
This root cause explains the vast majority (67 of 76) of bugs with the symptom of \textit{Resource Not Found}. 
In addition, as template engines evolve, breaking changes across versions can introduce errors. 
We identify 16 (1.59\%) bugs caused by \textit{Incorrect Template Engine Version}, such as the \textit{``removal of dynamic quotes functionality''} in Haml 5.0~\id{Haml}{60511890}. 
Finally, 9 (0.90\%) bugs arise from \textit{Improper Format Configuration} for character encoding, locales, or numeric formats, leading to discrepancies between rendered results and expected results. 

\textbf{Mechanism Misconception.} 
This category comprises 81 (8.07\%) bugs stemming from developer misconceptions regarding the engine's internal operation logic. 
Specifically, 45 (4.48\%) bugs arise from a lack of familiarity with engine-specific grammar rules (i.e., \textit{Language Convention}), such as operator precedence and variable scoping. 
Unlike \textit{Syntax Misuse}, template expression in these bugs are syntactically valid and pass compilation, but they produce unintended results during rendering. 
36 (3.59\%) bugs arise from \textit{Rendering Semantic}, i.e., how the engine transforms host-injected data into output. 
Specifically, 15 (1.49\%) bugs involve auto-escaping mechanisms designed to prevent Cross-Site Scripting (XSS) attacks. 
For example, while Jinja disables auto-escaping by default to support non-HTML rendering, Flask enables it automatically upon integration, leading to the unexpected rendering result in bug~\id{Jinja}{62984099}. 
Another 14 (1.39\%) bugs stem from a failure to respect the sequential nature of TE application lifecycle. 
As shown in Figure~\ref{fig: development-worflow}, there is a strict temporal separation between \textit{Template Rendering} and \textit{Rendered Result Processing}. 
This unidirectional flow prohibits the target consumer from updating template placeholder values during rendering stage~\id{Liquid}{78339503} and prevents the engine from accessing variables that only emerge during target processing stage~\id{Pug}{70050199}. 
Seven bugs involve the generation of stale artifacts due to the engine's IR caching. 

\textbf{Comparison with Cross-Language Bug Root Causes.} 
Figure~\ref{fig: root-causes} differentiates root causes shared with CLB~\cite{CLB} (in dotted borders) and those unique to TE applications (in solid borders). 
Our taxonomy shares only three root causes with~\cite{CLB}. 
The remaining 14 root causes are unique to TE applications and account for 581 (57.87\%) of the analyzed bugs. 
This significant divergence underscores that the failure modes of TE applications are fundamentally different from those of FFI-based multilingual software. 
Specifically, the unique characteristics of TE applications discussed in Section~\ref{s: intro}, such as \textit{Multi-Language Composition}, \textit{Opaque Data Flow}, and \textit{Complex Integration}, incur unique root causes related to syntax, data exchange and usage, and integration. 

\begin{figure}
    \centering
    \includegraphics[width=0.6\linewidth]{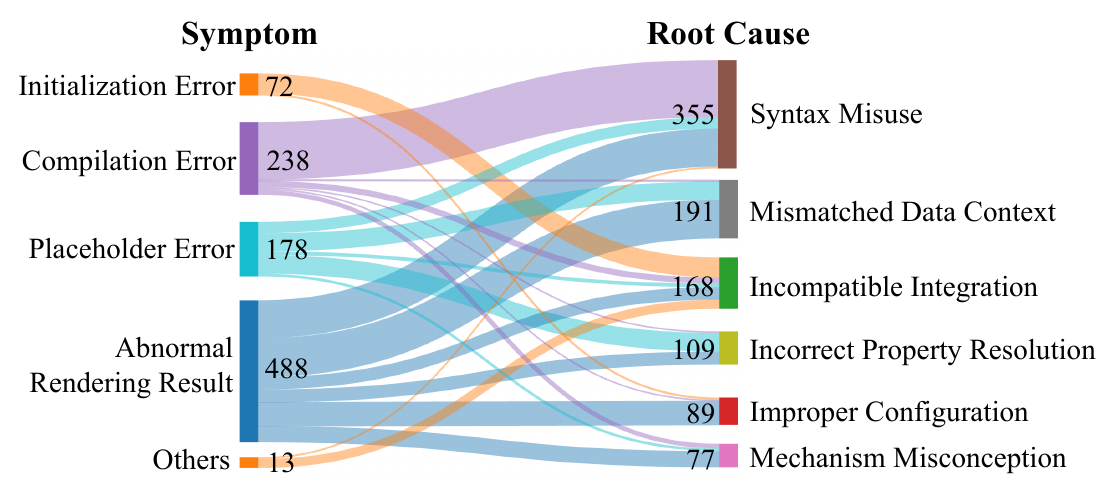}
    \caption{The relationship between symptoms and root causes.}
    \label{fig: symptom-root}
\end{figure}

\textbf{Relationship between Symptoms and Root Causes.} 
Similar to prior work~\cite{DLStack, Docker2025, DataVisualization2025}, we analyzed the relationship between observable symptoms and their underlying root causes, aiming to provide insights on the diagnosis of certain symptoms. 
We count the number of bugs for each symptom and root cause pair and visualize them in the Sankey diagram shown in Figure~\ref{fig: symptom-root}. 
Following prior work~\cite{DataVisualization2025}, we excluded pairs with fewer than five occurrences (representing 15 total bugs) to highlight the most significant patterns. 

Our analysis reveals three primary insights. 
First, most root causes exhibit predictable symptoms. 
For example, \textit{Syntax Misuse} primarily leads to \textit{Compilation Error} (188/355 = 52.95\%) or \textit{Abnormal Rendering Result} (124/355 = 34.93\%), while \textit{Mechanisms Misconception} leads predominantly to \textit{Abnormal Rendering Result} (53/77 = 68.83\%). 
Second, certain symptoms possess high diagnostic specificity for specific root causes. 
Notably, 78.99\% (188/238) of \textit{Compilation Error} bugs stem from \textit{Syntax Misuse}, and 90.28\% (65/72) of \textit{Initialization Error} bugs are caused by \textit{Improper Configuration}. 
This high concentration of causality suggests that developers can perform highly targeted triage when faced with such symptoms. 
Third, symptoms including \textit{Abnormal Rendering Result} and \textit{Placeholder Error} exhibit high causal pluralism, indicating significant debugging complexity. 
The root causes for \textit{Abnormal Rendering Result} are broadly distributed among \textit{Mismatched Data Context} (127/465 = 27.31\%), \textit{Syntax Misuse} (124/465 = 26.67\%), and \textit{Improper Configuration} (77/465 = 16.56\%). 
This explains the difficulty in resolving these bugs: unlike a compilation error, a rendering anomaly provides few immediate clues regarding whether the bug resides in the host-side data preparation, the template-side logic, or the environmental configuration. 
Similarly, \textit{Placeholder Error} bugs originate from a mix of \textit{Incorrect Property Resolution} (61/178 = 34.27\%), \textit{Mismatched Data Context} (59/178 = 33.15\%), and \textit{Syntax Misuse} (37/178 = 20.79\%). 

Finally, Figure~\ref{fig: symptom-root} reinforces the distinction boundaries between root cause categories. 
\textit{Syntax Misuse} is uniquely characterized by its high correlation with the early-stage \textit{Compilation Error} symptom. 
In contrast, \textit{Incorrect Property Resolution} and \textit{Mechanism Misconception} primarily trigger late-stage failures such as \textit{Placeholder Error} and \textit{Abnormal Rendering Result}, respectively. 
This divergence confirms that these categories represent fundamentally different types of defects. 

\begin{figure}
    \centering
    \includegraphics[width=0.6\linewidth]{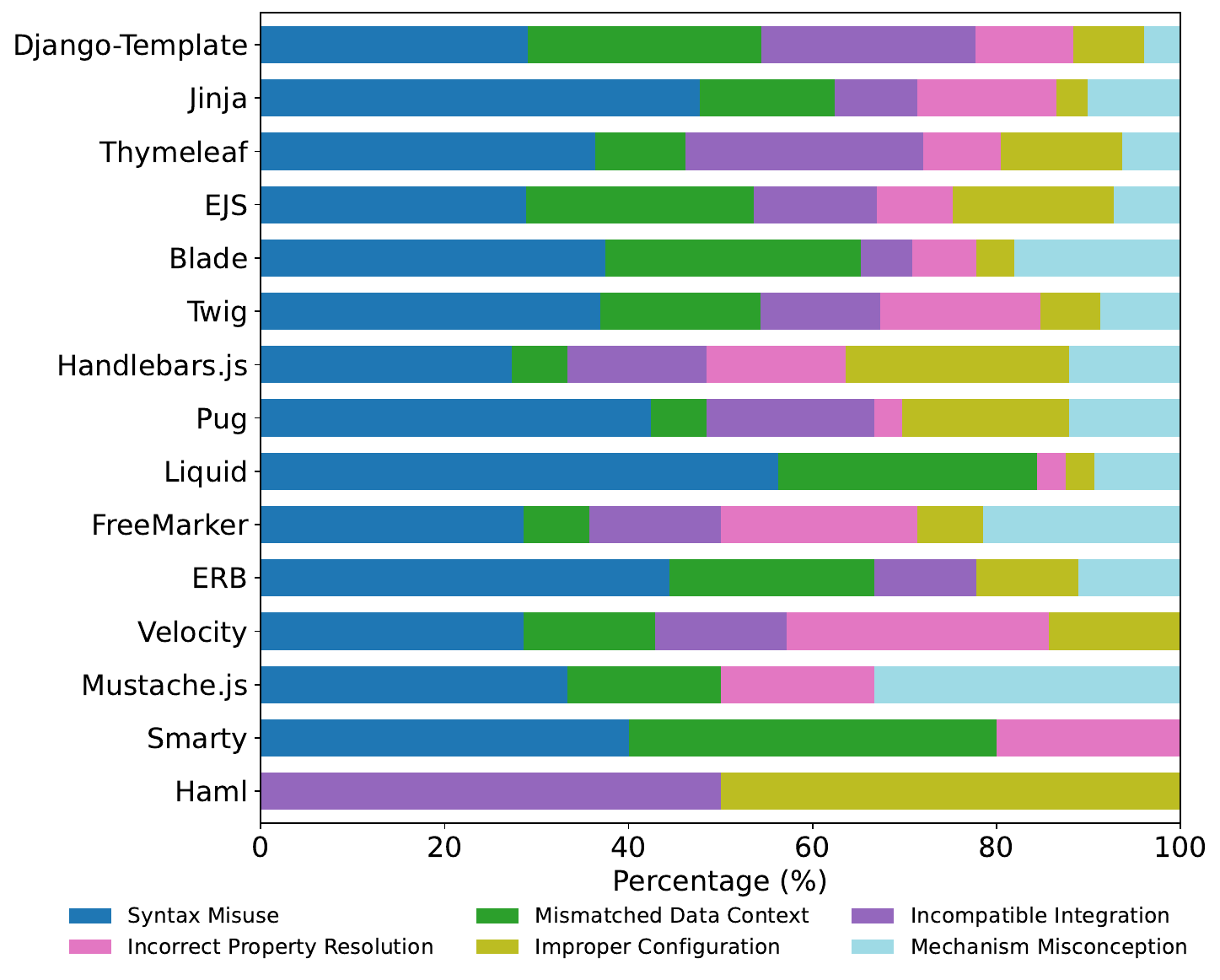}
    \caption{The root cause distribution of TE application bugs by engine.}
    \label{fig: te-root-cause}
\end{figure}

\textbf{Distribution of Root Causes across Template Engines.} 
Figure~\ref{fig: te-root-cause} presents the root cause distribution of TE application bugs for each of the 15 studied template engines. 
We observe that \textit{Syntax Misuse} remains the primary root cause for most engines, accounting for 27.27\% to 56.25\% of the bugs. 
This underscores an urgent need for specialized development tools that can assist practitioners in navigating the idiosyncratic syntax of templating languages. 
Similarly, \textit{Mismatched Data Context} is another prevalent root cause across nearly all engines, providing empirical evidence for the common opaque data flow challenge in TE application development regardless of the specific host environment or engine implementation. 
Furthermore, Figure~\ref{fig: te-root-cause} highlights the impact of architectural coupling: engines tightly integrated with web frameworks such as Django-Template (23.24\%) and Thymeleaf (25.87\%) exhibit a markedly higher incidence of \textit{Incompatible Integration} bugs, compared to standalone engines like Jinja (8.99\%). 
We performed a Kruskal-Wallis H-test on the distributions. The result indicates that the root cause distributions across the 15 engines exhibit no statistically significant difference ($p > 0.05$), suggesting that our taxonomy captures the fundamental, intrinsic bug characteristics inherent to TE applications. 

\begin{findingbox}
\textbf{Summary for RQ2:} TE application bugs stem from 17 root causes under six categories. 
\textit{Syntax Misuse}, \textit{Mismatched Data Context}, and \textit{Incompatible Integration} are the three most frequent root cause categories, collectively accounting for 71.91\% of the analyzed bugs. 
The root causes for \textit{Placeholder Error} and \textit{Abnormal Rendering Result} symptoms are diverse, underscoring the significant debugging complexity of these bugs. 
The root cause distributions across template engines are not significantly different.
\end{findingbox}

\subsection{RQ3: Fix Patterns}\label{ss: rq3-results}
\begin{figure}
    \centering
    \includegraphics[width=\linewidth]{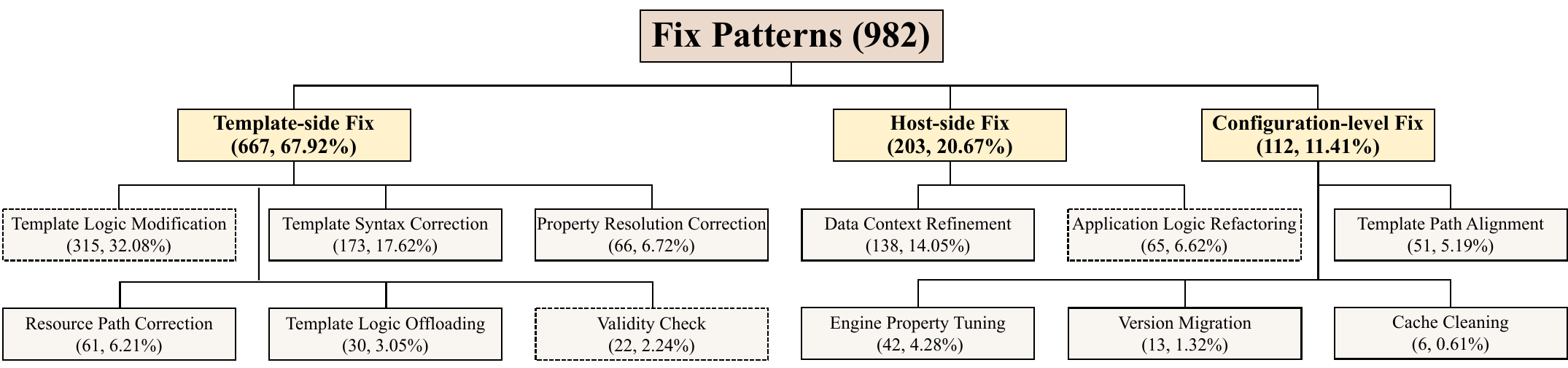}
    \caption{The taxonomy and distribution of fix patterns for TE application bugs. Dotted rectangles and solid rectangles represent fix patterns shared with CLBs and those unique to TE applications, respectively.}
    \label{fig: fix-patterns}
\end{figure}

Figure~\ref{fig: fix-patterns} illustrates the taxonomy and distribution of fix patterns for TE application bugs, which is structured into three high-level categories based on the fix site (i.e., \textit{Template-side Fix}, \textit{Host-side Fix}, and \textit{Configuration-level Fix}) and 12 leaf patterns. 
We are unable to infer fix patterns for 22 bugs due to insufficient information. 
So the analysis in this section is based on the remaining 982 bugs. 

\textbf{Template-side Fix.} 
The majority (667, 67.92\%) of fixes are implemented within the template. This category comprises six recurring patterns: 
\begin{itemize}[leftmargin=*, topsep=0pt]
    \item \textit{Template Logic Modification.} 
    This is the most common fix pattern in our dataset, where practitioners adjust the dynamic logic within templates to resolve runtime exceptions or achieve expected rendering results. 
    315 (32.08\%) of the analyzed bugs are fixed by this pattern. 
    These fixes frequently involves inline data transformation to ensure that host-injected values are compatible with the requirements of templating expressions or the target consumer. 
    For instance, a bug~\id{Pug}{67003953} was fixed by applying \Code{JSON.stringify()} to an array injected by the host, thereby enabling the correct consumption of client-side JavaScript within the rendered HTML document. 
    Other fixes involve utilizing specialized engine built-ins that developers had initially overlooked or re-implementing template logic to adhere to engine-specific semantic constraints. 
    \item \textit{Template Syntax Correction.} 
    This is the second most common fix pattern, accounting for 173 (17.62\%) of the analyzed bugs. 
    This pattern involves direct modifications to the template-side code to resolve syntax-level exceptions that halt the compilation or rendering process. 
    Examples include removing illegal nested delimiters or replacing the conflated host-language syntax with correct templating language syntax. 
    \item \textit{Property Resolution Correction.} 
    This pattern addresses erroneous property access from host-injected objects and fixes 66 (6.72\%) bugs. 
    Fixes typically involve either adopting engine-specific property access semantics, e.g., bug~\id{mustache.js}{71125277}, or ensuring alignment with properties explicitly exposed by the host's object model, e.g., bug~\id{Django-Template}{72771164}. 
    \item \textit{Resource Path Correction.} 
    To resolve \textit{Resource Not Found} bugs, developers update path expressions to adhere to the host framework’s resolution conventions, ensuring that static assets are locatable by the target consumer (e.g., the browser). 
    A frequent fix involves switching from template-relative paths to root-relative paths from the configured asset folder, e.g., \Code{/css/styles.css} in bug~\id{Handlebars.js}{59708405}. 
    This pattern was utilized to fix 61 (6.21\%) of the analyzed bugs. 
    \item \textit{Template Logic Offloading.} 
    This pattern involves offloading complex computation tasks from the template to the host environment, fixing 30 (3.05\%) of the analyzed bugs. 
    This pattern adheres to the common TE application development philosophy that templates should focus exclusively on presentation logic rather than complex business or program logic. 
    Developers typically implementing this by either pre-calculating values within the host-side code and injecting them as placeholder values, or by encapsulating the complexity within custom template filters. 
    \item \textit{Validity Check.} This pattern fixes 22 (2.24\%) bugs by adding conditional validity checks, e.g., testing for undefined or null before using a placeholder. 
\end{itemize}

\textbf{Host-side Fix.} 
This category encompasses fixes implemented within the host environment. 
It accounts for 203 (20.67\%) of the analyzed bugs and comprises two fix patterns. 
In particular, \textit{Data Context Refinement} involves adjustments to host-side data structures to ensure that they provide necessary values and expected data types required by the template logic. 
It is the third most common fix pattern, accounting for 138 (14.05\%) of the analyzed bugs. 
These fixes are typically localized and lightweight, such as explicitly serializing an object to a JSON string prior to data context injection in bug~\id{EJS}{63663849}. 
This pattern directly addresses the \textit{Mismatched Data Context} root cause. 
In contrast to simple context adjustments, \textit{Application Logic Refactoring} necessitates substantial modifications to the host's business logic, such as redesigning controllers or data models to facilitate data context preparation or to correctly handle requests issued by the target consumer. 
This pattern fixes 65 (6.62\%) of the analyzed bugs. 

\textbf{Configuration-level Fix.} 
This category consists of 112 (11.41\%) bugs fixed by modifying the environment settings of the template engine or the runtime environment. It comprises four fix patterns. 
Specifically, \textit{Template Path Alignment} directly addresses the \textit{Template File Path Misconfiguration} root cause. 
This pattern involves aligning the physical location of template files with the framework's automated discovery mechanisms. 
Developers either move templates to the framework's default search directory or explicitly override default configurations to specify custom paths. 
This pattern fixes 51 (5.19\%) bugs. 
42 (4.28\%) bugs are fixed by adjusting engine-specific parameters to ensure expected behaviors, i.e., \textit{Engine Property Tuning}. 
Common adjusted parameters include character encoding, auto-escaping policies, and locale. 
In addition, 13 (1.32\%) bugs are fixed by \textit{Version Migration}, i.e., upgrading or downgrading the engine, its associated plugins, or the host framework to a compatible version. 
Finally, 6 (0.61\%) bugs are fixed by \textit{Cache Cleaning} to prevent the generation of stale output. 

\textbf{Comparison with Cross-Language Bug Fix Patterns.} 
Figure~\ref{fig: fix-patterns} distinguishes fix patterns shared with CLB~\cite{CLB} (in dotted borders) from those unique to TE applications (in solid borders). 
Our taxonomy identifies nine fix patterns specific to TE applications, which collectively account for 580 (59.06\%) of the bugs. 
While three fix patterns, i.e., \textit{Template Logic Modification}, \textit{Application Logic Refactoring}, and \textit{Validity Check}, are similar to those identified in prior studies (e.g., \cite{CLB, Docker2025, DLStack}), the practical implementation can be fundamentally different due to the distinct nature of TE applications, such as limitations of the templating language and separation between business logic and presentation. 

\begin{figure}
    \centering
    \includegraphics[width=0.6\linewidth]{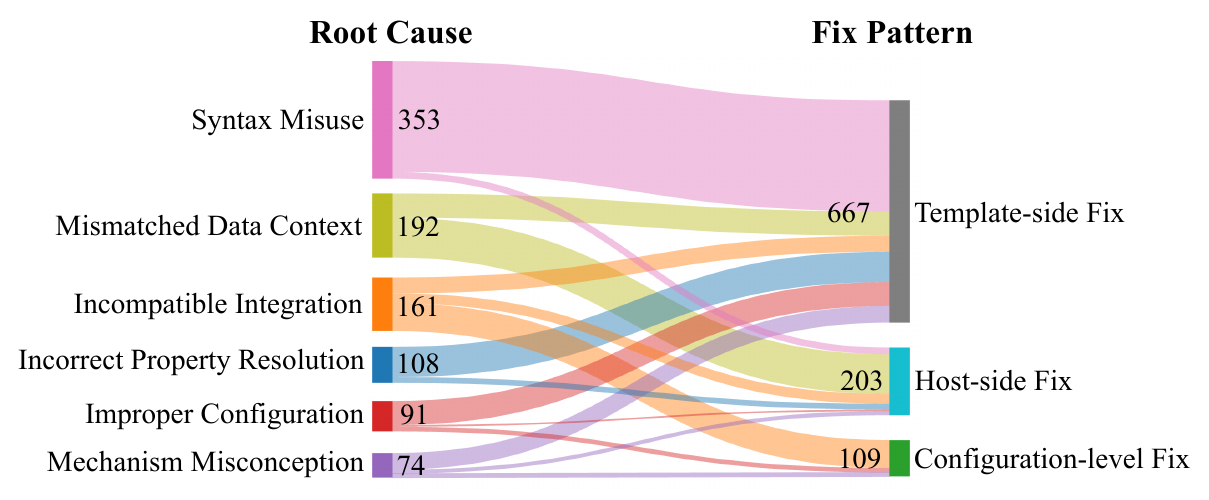}
    \caption{The relationship between root causes and fix patterns.}
    \label{fig: root-fix}
\end{figure}

\textbf{Relationship between Root Causes and Fix Patterns.} 
We analyzed the relationship between root causes and fix patterns, as shown in Figure~\ref{fig: root-fix}. 
Our analysis yields two primary insights. 
First, bugs stemming from template logic or syntax issues are predominantly fixed via template-side modifications. 
For root causes such as \textit{Syntax Misuse}, \textit{Incorrect Property Resolution}, \textit{Improper Configuration}, and \textit{Mechanism Misconceptions}, the fix typically resides within the template itself. 
This identifies the template as both the primary place for locating and repairing these bugs. 
Second, bugs arising from interactions across boundaries often necessitate cross-layer fixes. 
Specifically, although \textit{Mismatched Data Context} bugs manifest as template-level anomalies, 61.98\% (119/192) are fixed at the host-side, with the remainder being fixed within the template. 
This confirms that data-flow issues often necessitate changes to the host-side logic responsible for data context preparation. 
Furthermore, it reinforces the distinction between root cause \textit{Mismatched Data Context} (often requiring host-side fixes) and \textit{Incorrect Property Resolution} (predominantly fixed template-side). 
On the other hand, over half (83/161 = 51.55\%) of \textit{Incompatible Integration} bugs are fixed through configuration-level changes, while 29.81\% (48/161) require template-side fixes. 
This fix-site diversity underscores the inherent challenge of debugging TE applications: for approximately one-third of the analyzed bugs, identifying the root cause is merely the first step, as developers must then navigate across architectural boundaries, from template directives to host logic or environment settings, to implement a correct fix. 

\begin{figure}
    \centering
    \includegraphics[width=0.6\linewidth]{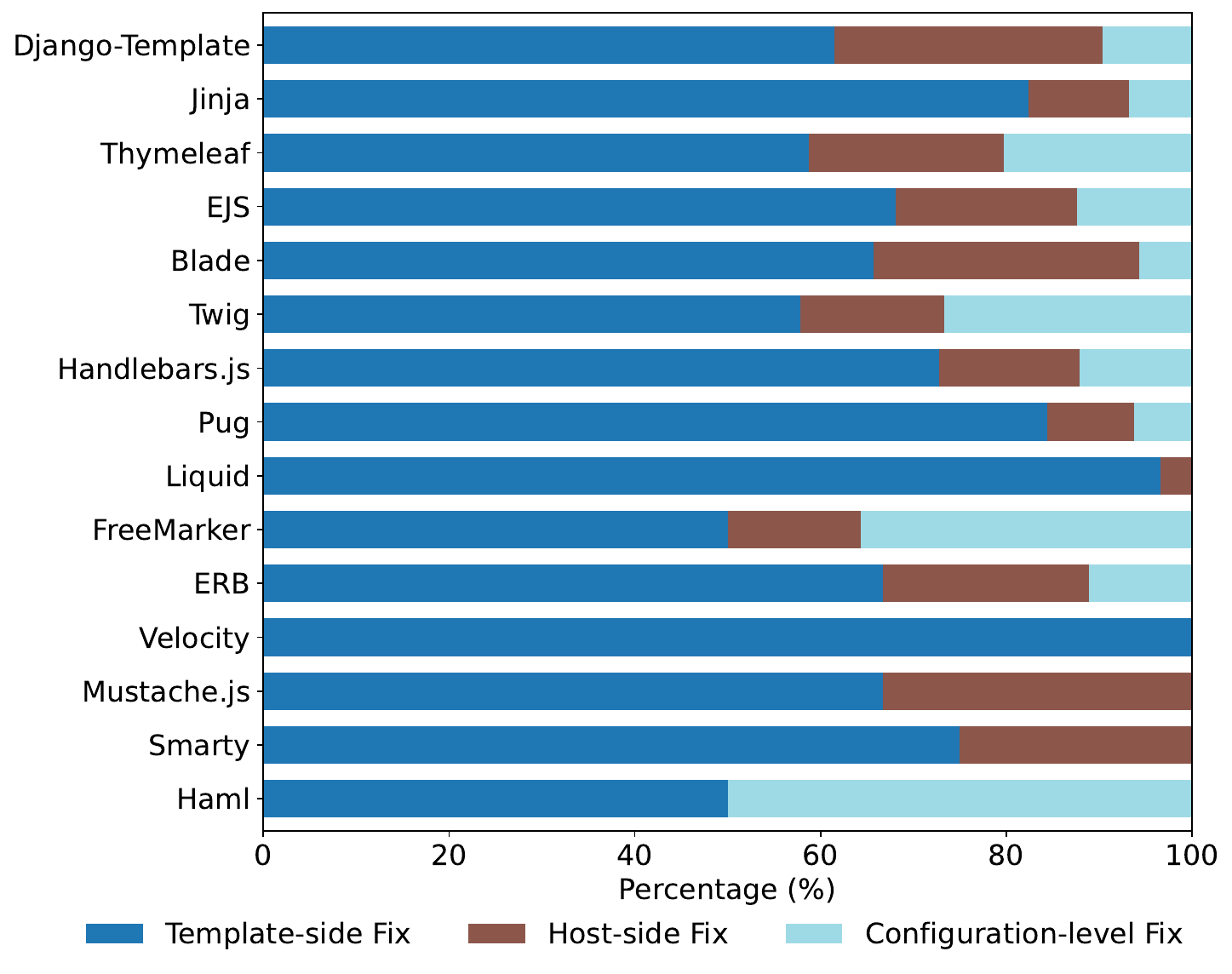}
    \caption{The fix pattern distribution of TE application bugs by engine.}
    \label{fig: te-fix-pattern}
\end{figure}

\textbf{Distribution of Fix Patterns across Template Engines.} 
Figure~\ref{fig: te-fix-pattern} presents the distribution of fix patterns for each of the 15 studied engines. 
We observe that \textit{Template-side Fix} is the dominant fix pattern across all the engines. 
This dominance is particularly pronounced in Jinja applications (82.39\%). 
This trend is likely driven by Jinja's widespread use in non-Web domains, such as Infrastructure-as-Code and data engineering. 
In these domains, templates are usually self-contained where the data context is defined within the template, thereby concentrating fix efforts within the template. 
The consistent prevalence of template-side fixes across all engines provides a strong empirical justification for the development of specialized repair tools for templates. 
In contrast, engines built into web frameworks, such as Django-Template (28.88\%) and Blade (28.57\%), exhibit a higher proportion of host-side fixes compared to standalone engines like Jinja (10.80\%). 
This reflects the architectural complexity of MVC-based applications, where the implicit coupling between host-side controllers and presentation layers often necessitates cross-boundary fixes. 
Furthermore, we observe that Twig (26.67\%) and Thymeleaf (20.29\%) have a higher proportion of configuration-level fixes, suggesting that practitioners should pay more attention to environment settings when using these engines. 
Overall, the fix pattern distributions across the 15 engines do not show statistically significant difference, as evidenced by a Kruskal-Wallis H-test ($p > 0.05$). 

\begin{findingbox}
\textbf{Summary for RQ3:} 
We identify 12 fix patterns for TE application bugs. 
While the majority (67.92\%) of fixes are implemented template-side, nearly one-third of bugs necessitate host-side (20.67\%) or configuration-level (11.41\%) fixes. 
Most root causes are fixed within the template, with exceptions of \textit{Mismatched Data Context} and \textit{Incompatible Integration}, which primarily require fixes in the host code and configuration settings respectively. 
Overall, the distribution of fix patterns does not show significant differences across the 15 studied engines. 
\end{findingbox}

\subsection{Taxonomy Validation on Bugs in GitHub Repositories}\label{ss: taxonomy_validation}
To ensure that the three taxonomies derived from SO reflect common bug characteristics of real-world TE applications, we further collected bugs from GitHub repositories, validating if they introduce new symptoms, root causes, or fix patterns that can not be covered by our taxonomies. 
Following prior work~\cite{Tensorflow2018, DeepLearing2019, DLStack, JupyterNotebbokBug}, we took two steps to collect TE application bugs from GitHub. 

\textbf{Step 1: GitHub Repository Selection.} 
We used GitHub Search API to obtain an initial set of candidate repositories based on three criteria: 
(i) the repository is written in the engine's corresponding host language; 
(ii) the engine's name appears in the repository's name, description, or README; 
(iii) the repository is not a fork. 
For example, to identify Thymeleaf-related repositories, we used the query: \Code{q=thymeleaf+language:java\&fork:false}. 
This phase yielded 26,252 candidate repositories. 
To isolate genuine TE applications, we then excluded repositories with empty or non-English descriptions and filtered out pedagogical repositories (e.g., tutorials, courses, and book examples) using established keywords~\cite{JupyterNotebbokBug}. 
This reduced the set to 19,767 repositories. 
Finally, to ensure the quality of the subjects, we retained only repositories with at least 100 commits, 10 stars, and 10 forks. 
We further verified the presence of template files using engine-specific file extensions (e.g., \Code{.html}, \Code{.j2}), resulting in a final set of 3,291 repositories.   

\textbf{Step2: Bug Collection.} 
We collected potential bugs from both issues and commits in the filtered repositories. 
To maintain consistency with our SO dataset, we applied the following inclusion criteria: 
(1) created after January 1, 2020; 
(2) contains bug-related keywords and template-specific terms (e.g., \textit{template}, \textit{render}); 
(3) for commits, they must involve a modification to either a template file or a host-side code file; 
(4) written in English. 
This process yielded 1,465 candidate issues and 4,548 candidate commits. 
As the data volume is still large, we randomly sampled 600 issues and 600 commits, a size significantly exceeding the statistically representative requirements (314 and 365 respectively) for a 95\% confidence level with a 5\% margin of error. 
We then performed a manual review to identify TE application bugs, following the protocol established in Section~\ref{ss: bug_collection}. 
The manual labeling yielded a Cohen’s Kappa of 0.91, and the conflicts were resolved with a discussion. 
Ultimately, we identified 180 bugs, 93 from issues and 87 from commits. 

The two authors then independently mapped the 180 GitHub bugs to the SO-derived taxonomies, achieving high inter-rater reliability with the Cohen’s Kappa coefficient of 0.94, 0.89, and 0.84 for symptoms, root causes, and fix patterns respectively. 
We found that the 180 GitHub bugs' symptoms, root causes, and fix patterns are all covered by our taxonomies. 
We employed Mann-Whitney U test~\cite{mann-Whitney} with Holm-Bonferroni correction~\cite{Holm-Bonferroni} to compare the distribution of bugs between the two platforms. 
The results indicate no statistically significant difference in the distribution of symptoms and root causes between the two datasets. 
While the distribution of fix patterns shows a statistically significant difference ($p < 0.01$), the effect size was negligible (0.1044), suggesting high cross-platform stability. 

Consistent with SO data, \textit{Abnormal Rendering Result} remains the most common symptom on GitHub (83, 48.61\%). 
However, we observe several distribution shifts on GitHub data: 
GitHub bugs exhibit a higher proportion of \textit{Initialization Error} symptom (22.22\% vs. 7.67\% on SO) and \textit{Incompatible Integration} root cause (33.89\% vs. 16.73\%). 
Conversely, GitHub bugs exhibits a lower frequency of \textit{Compilation Error} symptom (13.89\% vs. 23.71\%) and \textit{Syntax Misuse} root cause (21.11\% vs. 35.76\%). 
Consequently, GitHub bugs require more host-side fixes (27.78\% vs. 20.67\%) and configuration-level fixes (20.00\% vs. 11.41\%) than SO, though \textit{Template-side Fix} remains the dominant strategy (52.22\%). 
These shifts align with the distinct development contexts of the two platforms. 
SO hosts many syntax-related questions from novices when they experiment with template engines in isolation. 
In contrast, template engines are typically integrated with external frameworks in GitHub repositories, inevitably leading to a higher frequency of integration and environment bugs. 
Consistent with~\cite{Tensorflow2018}, we consider that syntax-related bugs are of equal significance, as they highlight the deficiency of fundamental development support in the template engine ecosystem. 
In summary, this validation study confirms that the taxonomies derived from SO data are robust and offer a comprehensive characterization of real-world TE application bugs, spanning scenarios from standalone usage to complex integrations. 

\section{Implications and Prototype Tools}
\subsection{Implications}
\subsubsection{Implication for Tool Designers}\label{sss: too designers} 
Our study reveals a significant gap between the architectural complexity of TE applications and the capabilities of existing development tools. 
We surveyed top-ranked VS Code extensions for the studied template engines (e.g., Django, Better Jinja, and EJS Language Support) and found that current support is largely rudimentary, primarily limited to basic syntax highlighting and snippets. 
Such features are insufficient to address the diverse and complex bugs identified in our study. 
Based on our findings, we propose three critical functionalities for next-generation TE application development tools: 

(1) \underline{\textit{Syntax-Aware Error Detection and Automated Repair.}} 
We find that \textit{Compilation Error} is the second most common (23.71\%) symptom of TE application bugs (RQ1), underscoring the prevalent issue of correctly using templating language syntax. 
Therefore, we argue that support tools must evolve beyond highlighting to provide active syntax validation and automated fixes for common errors, such as nested or mismatched delimiters. 
To demonstrate this, we develop a prototype tool for Jinja in Section~\ref{sss: syntax tools} that targets four types of syntax errors. 

(2) \underline{\textit{Cross-Layer Semantic Auto-completion.}} 
Our study reveals that a considerable portion (18.03\%) of TE applications bugs manifest as \textit{Placeholder Error} (RQ1) and primarily arise from \textit{Mismatched Data Context} or \textit{Incorrect Property Resolution} (RQ2). 
These bugs are a direct consequence of the opaque data flow between the host and the template as introduced in Section~\ref{s: intro}. 
Because template composition and business logic preparation often occur in parallel, there is no formal contract ensuring that the host provides the data the template expects and the template correctly uses the data as the host provides. 
To mitigate this, tools should bridge the host-template boundary by: 
(i) extracting placeholder requirements (e.g., names, data types, and properties) from templates to provide hints during host-side data context preparation, 
and (ii) resolving host-side data structures to enable placeholder and property auto-completion during template composition. 
While dbt Labs recently proposes \textit{TypeJinja}~\cite{TypeJinja} that detects type errors in dbt Jinja templates, it assumes a self-contained environment where the data context is contained in the template. 
In contrast, our findings suggest that tools must handle external data contexts originating from separate host-side files. 
This functionality requires a template element extraction utility such as the one we present in Section~\ref{sss: extractor}. 

(3) \underline{\textit{Integrated Consumer-Side Preview with Mock Injection.}} 
Our study also reveals that many TE application bugs remain latent until processed by the target consumer, e.g., \textit{Resource Not Found} and \textit{Broken HTML Elements} (RQ1). 
This reflects the deferred validation characteristic of TE applications discussed in Section~\ref{s: intro}. 
To alleviate this, tools should provide real-time, WYSIWYG (What You See Is What You Get) previews that simulate the target consumer's environment. 
Furthermore, this preview should support mock data injection, allowing developers to provide sample placeholder values and observe how dynamic content transformations behave without requiring a full application deployment. 

\subsubsection{Implication for Practitioners} 
Based on the root causes and fix patterns identified in our study, we derive several \textbf{template composition guidelines} for practitioners: 

(1) \underline{\textit{Adhere to the Principle of Minimal Template Logic.}} 
Consistent with the design philosophy of many template engines, templating languages are specialized for dynamic content generation rather than general-purpose programming. 
These domain-specific templating languages often lack the expressive power and debugging support as general-purpose languages. 
However, our study identifies \textit{Expression Misuse} as a dominant root cause for TE application bugs (RQ2), primarily stemming from attempts to implement complex logic within the template. 
Furthermore, we find that 3.05\% of bugs are fixed by offloading complex computations to the host (RQ3). 
Therefore, practitioners should minimize business logic within templates and focus exclusively on presentation. 
Complex data processing should be performed within the host environment, with results passed as placeholder values, or encapsulated within custom template tags and filters to maintain a clean separation of concerns. 

(2)\underline{\textit{Navigate Polyglot Syntactic Boundaries Diligently.}} 
As introduced in Section~\ref{s: intro}, TE application development typically involves navigating at least three languages, i.e., the host language (e.g., Python), the templating language (e.g., Jinja), and the target language (e.g., HTML). 
Besides, the template file involves both the templating language and the target language. 
Without a clear understanding of the syntactic boundaries and potential conflicts between these languages, developers are prone to syntactic conflation, leading to compilation failures or abnormal rendering result. 
This is evidenced by the 3.39\% of bugs caused by \textit{Syntax Confusion} (RQ2), where practitioners incorrectly attempted to utilize host-specific modules or target-specific comment styles within the templates. 

(3) \underline{\textit{Respect the Temporal Separation of the Execution Lifecycle.}} 
As illustrated in Figure~\ref{fig: development-worflow}, TE applications follow a strict, sequential execution lifecycle: the host first renders the template, and only then is the resulting document processed by the target consumer (e.g., a browser). 
Violating this temporal separation, e.g., attempting to update template placeholder values via target-side JavaScript, definitely leads to bugs, as exemplified by bugs in \textit{Rendering Semantic} (RQ2). 
To ensure correct data exchange, developers must adhere to the unidirectional data flow: data must be captured from the target consumer (e.g., via HTTP requests), processed by the host environment, and then injected into the template engine to generate the updated view. 

In addition to the template composition guidelines, our study also provides systematic \textbf{bug localization and fix guidance} for practitioners, based on the analysis of relationships between symptoms, root causes, and fix patterns. 
On the one hand, the relationships analyzed in RQ2 enable practitioners to prioritize specific diagnostic paths based on observable symptoms. 
First, we find 90.28\% of \textit{Initialization Error} symptoms stem from \textit{Incompatible Integration}. 
Therefore, practitioners should first check their environment settings. 
Specifically, for \textit{Template Not Found} errors, developers should verify the alignment between the actual template location and the framework’s template discovery paths. 
Second, when encountering \textit{Undefined Variable} or \textit{Type Mismatch} symptoms, the high correlation with \textit{Mismatched Data Context} suggests that practitioners should first inspect the host-side logic responsible for context preparation.
Third, for \textit{Property Access Errors}, diagnostic efforts should be directed toward the template's resolution syntax and the host's object model. 
On the other hand, the relationship between root causes and fix patterns analyzed in RQ3 can shed light on effective bug fixes. 
For example, when a bug is traced to \textit{Incorrect Property Resolution}, our findings suggest that the fix is typically achieved through template-side syntax or property name correction. 
Conversely, for bugs rooted in \textit{Mismatched Data Context}, practitioners should be prepared for host-side fixes, as over 60\% of such bugs require modifying the host logic. 

\subsubsection{Implication for Researchers} 
Our findings open several promising avenues for future research into the reliability and quality assurance of TE applications:

(1) \underline{\textit{Cross-Layer Data Flow Analysis.}} 
TE applications involve implicit data exchange across the host, template, and target consumer, which results in distinct bug characteristics from FFI-based multilingual software, such as \textit{Mismatched Data Context}, \textit{Incorrect Property Resolution}, and \textit{Incorrect Requests}. 
Future research could explore data flow analysis techniques across the host-template boundary. 
Such techniques could help formalize the ``implicit contracts'' between these layers and enable more robust static verification of data integrity throughout the application lifecycle. 

(2) \underline{\textit{Multimodal Automated Program Repair.}} 
Diagnosing \textit{Abnormal Rendering Result} (RQ1) often requires a holistic view of the host code, the template directives, and the visual output (e.g., screenshots of the rendered HTML). 
The emergence of Multimodal Large Language Models (MLLMs), which can simultaneously process both texts and images, presents a unique opportunity. 
Researchers could explore the potential of MLLMs to automate this cognitively demanding process. 

\subsection{Prototype Tools}
To demonstrate the practical utility of our findings, we develop two prototype tools for Jinja templates: (1) a Syntax Error Detection and Repair tool and (2) a Semantic Element Extractor. We selected Jinja as our target engine for three primary reasons: 
(1) \textit{Independence}. Unlike engines such as Blade or Django-Template, which are tightly coupled with specific web frameworks, Jinja is a standalone engine that can be integrated into diverse environments. 
(2) \textit{Versatility}. It is widely adopted across multiple domains, including web development, Infrastructure-as-Code, and data engineering. 
(3) \textit{Representativeness}. It constitutes the second most bugs in our dataset, providing rich data for tool validation. 

\begin{table}
    \centering
    \caption{Detection and Refinement Rules for Jinja2 Template Syntax Errors.}
    \label{tab:jinja_rules}
    \footnotesize
    \renewcommand{\arraystretch}{0.95}
    \setlength{\tabcolsep}{4.5pt}
    \begin{tabularx}{\columnwidth}{@{} p{2.0cm} >{\raggedright\arraybackslash}X >{\raggedright\arraybackslash}X @{}}        
    \toprule
        \textbf{Error} & \textbf{Detection} & \textbf{Refine Strategy} \\
        \midrule
        Nested \newline Delimiters & 
        Identify nested \texttt{\{\{ \}\}} inside other delimiters (e.g., inside \texttt{\{\% \%\}}). 
        \textbf{e.g.} \texttt{\{\% if \{\{user\}\} \%\}} & 
        Remove the redundant internal brackets and use the variable name directly. 
        \textbf{Fix:} \texttt{\{\% if user \%\}} \\
        \addlinespace[3pt]
        
        Misplaced \newline Extends Tag & 
        Identify \texttt{\{\% extends ... \%\}} tags that do not appear at the very beginning of the file. \par
        \textbf{e.g.} \texttt{\{\{item\}\}\{\% extends.. \%\}} & 
        Relocate the \texttt{extends} tag to the top of the template file to ensure proper inheritance. \par
        \textbf{Fix:} \texttt{\{\% extends.. \%\}\{\{item\}\}} \\
        \addlinespace[3pt]
        
        Mismatched \newline Delimiters & 
        Unclosed or mismatched delimiters (e.g., \texttt{\{\% \}\}}) and unbalanced block tags. \par
        \textbf{e.g.} \texttt{\{\% if user} & 
        Complete and match the delimiters (\texttt{\{\% \%\}}, \texttt{\{\{ \}\}}) and tags (e.g., \texttt{if/endif}). \par
        \textbf{Fix:} \texttt{\{\% if user \%\}} \\
        \addlinespace[3pt]
        
        Invalid \newline Property Access & 
        Identify unsupported accessor symbols (e.g., \texttt{->}) or invalid formats. \par
        \textbf{e.g.} \texttt{user->name} & 
        Convert the access syntax to the standard Jinja format using dot (\texttt{.}) or bracket notation (\texttt{[]}).\par
        \textbf{Fix:} \texttt{user.name} \\
        \bottomrule
    \end{tabularx}
\end{table}

\subsubsection{Jinja Syntax Error Detection and Repair Tool}\label{sss: syntax tools}
Our study reveals that \textit{Compilation Error} (23.71\% in RQ1) is a significant hurdle, yet existing development tools (e.g., Better Jinja) fail to provide active detection and fix. 
To address this, we develop a rule-based prototype tool capable of automatically detecting and fixing common syntax errors.

Specifically, we analyze the 33 Jinja application bugs in our dataset that are fixed via \textit{Template Syntax Correction} (RQ3). 
We filtered for bugs that could be statically detected and whose fix involves only Jinja syntax, resulting in 10 bugs. 
By analyzing their root causes and fixes, we derive two error classes: \textit{Nested Delimiters} and \textit{Misplaced Extends Tag}. 
We further enrich these with two additional error classes based on Jinja syntax documentation: \textit{Mismatched Delimiters} and \textit{Invalid Property Access Syntax}. 
The resulting detection and refinement strategies for them are summarized in Table~\ref{tab:jinja_rules}. 
We implement the prototype tool based on these rules. 

To evaluate the effectiveness of the tool across diverse scenarios, we curate a test suite of 30 bugs, which includes 10 real-world bugs from our filtered dataset and 20 synthetically augmented bugs designed to test cases not covered by the 10 real-world bugs. 
Specifically, for the first error class, in addition to the nine real-world bugs, we synthesize two bugs by nesting \Code{\{\{ \}\}} inside \Code{\{\{ \}\}} and \Code{\{\% \%\}} respectively. 
For the second error class, in addition to one real-world bug, we synthesize three bugs by varying the position of the \Code{extends} tag. 
For the third error class, we synthesize 10 bugs by removing close delimiters and mismatching delimiters. 
For the last error class, we synthesize 5 bugs using various combinations of invalid access, such as \Code{user->name} and \Code{user[name]->email}. 
We manually inspect the tool’s output for each bug to verify both detection accuracy and the correctness of the proposed fix. 
Our tool successfully detects the syntax errors and generates correct fixes for all 30 bugs, demonstrating its effectiveness. 

\subsubsection{Jinja Template Element Extractor}\label{sss: extractor}
As revealed in RQ1 and RQ2, the absence of an explicit placeholder contract between the template and the host leads to a high frequency of \textit{Undefined Variable}, \textit{Property Access Error}, and \textit{Type Mismatch} errors. 
To mitigate this opaque data flow issue, development tools should bridge the data gap between the host and template. 
As a step towards this goal, we develop a static analysis-based prototype tool that extracts the schema requirements of a template, including placeholders, tags, and filters. 

Given a Jinja template, the extractor traverses its Abstract Syntax Tree (AST) to generate a hierarchical representation of the required data context. 
The tool operates as follows. 
It identifies all global variables required by the template. It explicitly filters out local variables (e.g., those defined within \Code{macro} blocks, \Code{set} assignments, or as loop iterators) as they do not impose requirements on the host-side data context. 
Upon encountering a property access node (e.g., \Code{a.b} or \Code{a['b']}), the tool recursively resolves the object hierarchy in our placeholder tree and appends the accessed attribute to the parent object's schema. 
When a placeholder is utilized within an iterative construct (e.g., a \Code{for} loop), the tool infers that the variable must be an \textit{Iterable} type. 
It then treats the loop variable as a child of the placeholder and captures its information, similar to placeholders. 
The tool also catalogs all tags and filters, distinguishing between Jinja’s built-ins and custom extensions. 
This provides a checklist for developers to ensure that all required custom logic is correctly registered in the host environment. 

To evaluate the extractor’s effectiveness, we curate a test suite of 50 templates sampled from our dataset. 
To reflect Jinja’s broad application scenarios, this suite includes 20 HTML templates (Web), 20 YAML templates (Ansible), and 10 SQL templates (dbt). 
We specifically prioritize templates with high structural complexity, such as deeply nested loops and multi-stage filter chains. 
We manually verify the extracted schema against the template code to determine correctness. 
Our tool achieves an accuracy of 96\% (48 of 50 templates). 
The two failed cases result from over-approximation, where the tool incorrectly interprets built-in Python methods as dynamic data attributes. 
We plan to address this by integrating a comprehensive library of host language built-ins in the future. 

\section{Threats to Validity}
The major threat to internal validity is the potential of subjectivity during the manual labeling process. 
To mitigate this, we followed a rigorous open coding procedure involving three researchers with rich experience in TE application development. 
Furthermore, we developed s coding guide and measured inter-rater agreements using the Cohen’s Kappa coefficient $\kappa$. 
All $\kappa$ scores are greater than 0.8, indicating almost perfect agreement. 
Second, similar to prior work~\cite{DLSPB}, we derive the taxonomies based on bugs from SO, which may miss characteristics on other platforms such as GitHub. 
To mitigate this, on the one hand, we selected diverse (15) template engines across five programming languages and collected 1,004 bugs; 
on the other hand, we collected 180 bugs from GitHub repositories to validate the completeness of our taxonomies. 
The validation results confirms that our taxonomies are complete. 
Finally, our tools are prototypes. While they demonstrate high accuracy on our test suite, their performance may vary when integrated into diverse, high-complexity production environments. 
In the future, we will integrate them into a VS Code extension and evaluate them on real-world projects. 

\section{Related Work}
\textbf{Research on Template Engines.} 
Despite their ubiquity, research on template engines remains scarce. 
Given their wide adoption in web applications, existing literature focuses predominantly on security vulnerabilities, particularly Server-Side Template Injection (SSTI) attacks~\cite{Kettle2015SSTI}. 
Zhao et al.~\cite{Zhao2023USENIX} proposed \textsc{TEFuzz} to automatically detect and exploit template escape bugs that bypass engine sandboxes. 
Pisu et al.~\cite{2024Survey} conducted a comprehensive survey of practical SSTI vulnerabilities and detection tools. 
While vital, these studies do not investigate bugs in TE applications. 

Template engines also serve as key components of Infrastructure-as-Code (IaC) and data orchestration tools. 
Recent studies on these domains have begun to highlight the prevalence and complexity of template-related issues. 
Begoug et al.~\cite{IaC2023} and Carreira et al.~\cite{AnsibleChallenges} found that templating-related questions are among the most frequent and challenging topics discussed by IaC practitioners, while Drosos et al.~\cite{IaC2024} identified template bugs as a primary root cause of IaC program bugs. 
More recently, Ding et al.~\cite{TypeJinja} proposed \textsc{TypeJinja} to detect type-related errors within Jinja-based dbt workflows. 
However, these studies focus on domain-specific contexts and do not offer a generalized and fine-grained characterization of TE application bugs. 
Complementary to these efforts, our work presents the first large-scale study of TE application bugs based on 1,004 bugs across 15 engines, providing a comprehensive view of their symptoms, root causes, and fix patterns. 

\noindent \textbf{Research on Multilingual Software.} 
Our study is situated within the broader context of multilingual software development~\cite{mixedLangProg}, where multiple programming languages are integrated into a single software project. 
While this paradigm is prevalent in modern software engineering~\cite{Mayer2017, PolyglotGitHub, MPLOSS}, it introduces significant challenges, including complex cross-language interfacing~\cite{10413900, MLSDChallenge}, increased bug-proneness~\cite{7476675, MLDSmell, MLCVulnerability, JNIBug}, and high fix complexities for multi-language bugs~\cite{10304793, MLDLFBug}. 
Prior work has extensively investigated cross-language bugs in Foreign Function Interface-based multilingual software (FFI-MLS), such as Python-C~\cite{CLB, MLB, Insight, 10.1002/smr.2507}, Java-C~\cite{CLB, Insight}, and C-Fortran~\cite{CFortran}. 
Our work enriches the literature by investigating a distinct class of multilingual software, i.e., TE applications, which present significantly different bug characteristics from FFI-MLS as evidenced by the comparison in Section~\ref{s: results}. 

\noindent \textbf{Empirical Study on Bug Characteristics.} 
Understanding bug characteristics (e.g., symptoms, root causes, and fix patterns) is a prerequisite for designing specialized automation tools for bug detection, localization, and repair. 
This empirical approach has been successfully applied to a variety of domains, including 
\textit{Deep Learning} (frameworks~\cite{DeepFramework2023, FLSystemBugs, DLFrameworkBugs2, MLModelOptimizationBugs} and applications~\cite{Tensorflow2018, DeepLearing2019, Deployment2021, DLStack, DLSytemFaults, DLModelFix, DLSPB}), 
\textit{Data Science} (libraries~\cite{DataVisualization2025, DataScience2025} and Jupyter notebooks~\cite{JupyterNotebbokBug}), 
\textit{System Software} (compilers~\cite{DeepLearningCompiler, CPPCompilers}), autonomous driving systems~\cite{ADSBugs, ADSFix}, container runtime systems~\cite{Container2024, Docker2025} and orchestration tools~\cite{KubernetesOperatorBugs}), 
\textit{Mobile Applications}~\cite{AndroidAppBugs1, AndroidAppBugs2}, 
and GPU programs~\cite{GPU2025}. 
Despite this breadth of research, the template engine ecosystem remains as a ``black box''. 
In this paper, we fill this gap by presenting the first comprehensive bug characterization study on TE applications. 

\section{Conclusion}
This paper presents the first comprehensive study of TE application bugs. 
We systematically analyze 1,004 bugs across 15 template engines and five languages and construct taxonomies of symptoms, root causes, and fix patterns. 
Our findings reveal that \textit{Abnormal Rendering Result} is the most common symptom, primarily manifesting as silent failures such as unexpected or blank output. 
\textit{Syntax Misuse}, \textit{Mismatched Data Context}, and \textit{Incompatible Integration} are the frequent root causes of TE application bugs. 
While 67.92\% of the bugs are fixed within the template, nearly one-third require modifications to the host code or environment settings. 
Based on these findings, we provide actionable implications on development tool design, template composition guidelines, bug localization and fix guidance, and future research. 
We also develop two prototype tools for Jinja applications. 
In the future, we plan to integrate the two tools into a VS Code extension and incorporate more functionalities to facilitate TE application development. 

\begin{acks}
This work is supported by the National Natural Science Foundation of China under grants 62502030 and 92582119. 
\end{acks}

\bibliographystyle{ACM-Reference-Format}
\bibliography{references}

\end{document}